\documentclass[aps,prl,twocolumn,showpacs,superscriptaddress]{revtex4-1}%
\usepackage{graphicx}
\usepackage{color}
\usepackage{amsmath}
\usepackage{amsfonts}
\usepackage{amssymb}%
\providecommand{\U}[1]{\protect\rule{.1in}{.1in}}

\begin{document}
\title{Uncovering the mechanism of the impurity-selective Mott transition in
paramagnetic V$_{2}$O$_{3}$}
\author{Frank Lechermann}
\affiliation{I. Institut f\"ur Theoretische Physik, Universit\"at~Hamburg, 
Jungiusstr.~9, D-20355~Hamburg, Germany}
\author{Noam Bernstein}
\affiliation{Code 6393, Naval Research Laboratory, Washington, DC 20375, USA}
\author{I. I. Mazin}
\affiliation{Code 6393, Naval Research Laboratory, Washington, DC 20375, USA}
\author{Roser Valent{\'i}}
\affiliation{Institut f{\"u}r Theoretische Physik, Goethe-Universit\"at Frankfurt,
Max-von-Laue-Str. 1, 60438 Frankfurt am Main, Germany}

\begin{abstract}
While the phase diagrams of the one- and multi-orbital Hubbard model have been
well studied, the physics of real Mott insulators is often much richer,
material dependent, and poorly understood. In the prototype Mott insulator
V$_{2}$O$_{3}$, chemical pressure was initially believed to explain why the
paramagnetic-metal to antiferromagnetic-insulator transition temperature is
lowered by Ti doping while Cr doping strengthens correlations, eventually
rendering the high-temperature phase paramagnetic insulating. However, this
scenario has been recently shown both experimentally and theoretically to be
untenable. Based on full structural optimization, we demonstrate via the
charge self-consistent combination of density functional theory and dynamical
mean-field theory that changes in the V$_{2}$O$_{3}$ phase diagram are driven
by defect-induced local symmetry breakings resulting from dramatically different
couplings of Cr and Ti dopants to the host system. This finding emphasizes the
high sensitivity of the Mott metal-insulator transition to the local
environment and the importance of accurately accounting for the one-electron
Hamiltonian, since correlations crucially respond to it.

\end{abstract}

\maketitle

\begin{figure}[t]
\centering
\includegraphics*[width=8cm]{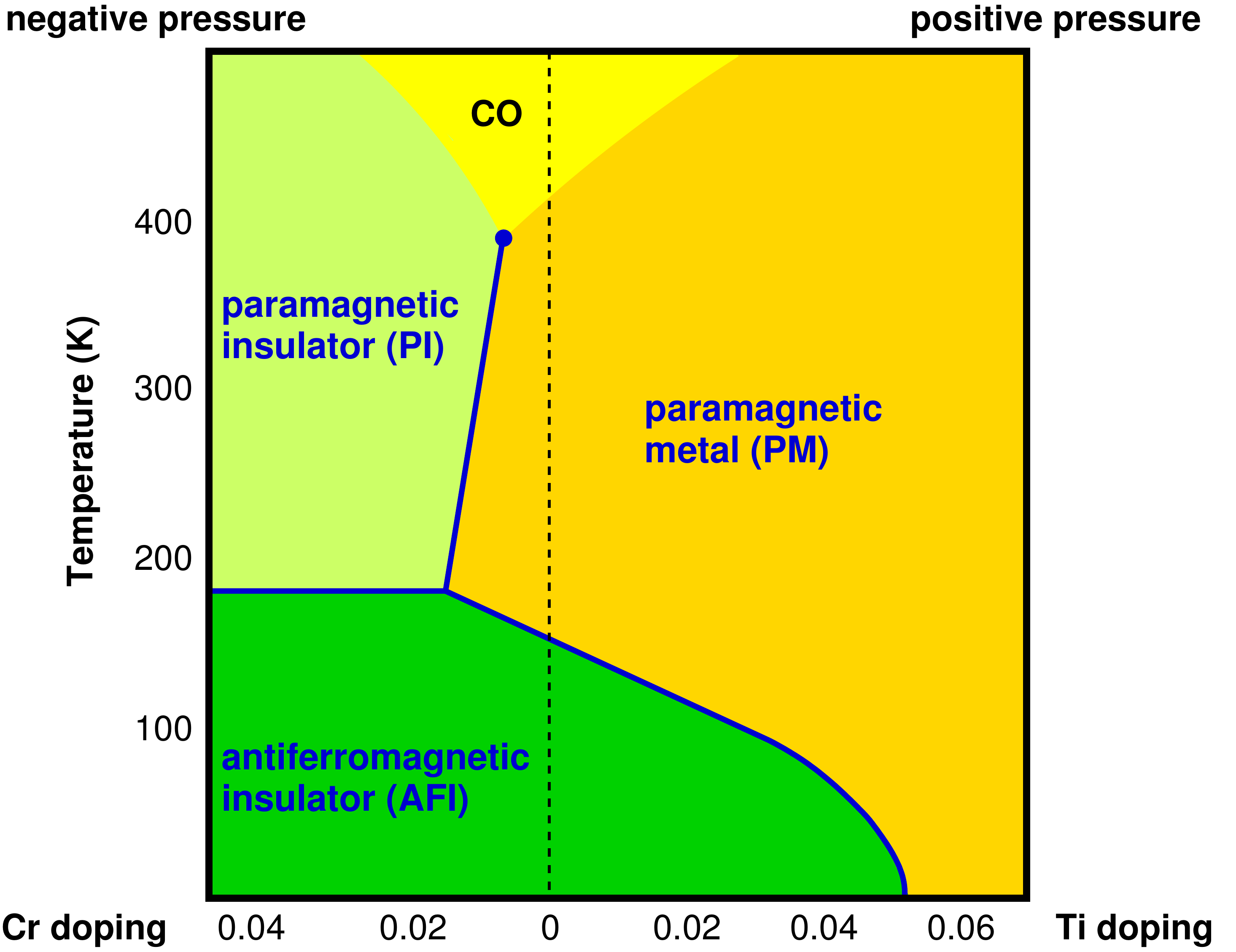}\caption{(Color online) V$_{2}$O$_{3}$
phase diagram (after~\cite{mcw69,*mcw71,*mcw73}), with 'CO' marking the
crossover region}%
\label{fig:pd}%
\end{figure}

\textit{Introduction.---} After many decades of intense research, the V$_{2}%
$O$_{3}$ phase diagram still stands out as a challenge in condensed matter
physics~\cite{ric70,cas78,par00,hel01,tan09,fuj11}. Its canonical form (cf.
Fig.~\ref{fig:pd}) has been addressed many times both experimentally and
theoretically~\cite{mcw69,*mcw71,*mcw73}. At stoichiometry and at ambient
temperature $T$, a paramagnetic-metallic (PM) phase in a corundum crystal
structure is stable. Driving the temperature below $T_{\mathrm{N}}\sim
155$\thinspace K results in an antiferromagnetic-insulating (AFI) phase with a
monoclinic crystal structure. Ti doping rapidly suppresses this low-$T$ phase,
whereas Cr doping stabilizes it, and, additionally, transforms the high-$T$ PM
phase into a paramagnetic-insulating (PI) one.

V$^{3+}$has two $3d$ electrons, occupying the triply degenerate (for each
spin) $t_{2g}=\{e_{g}^{\pi},a_{1g}\}$ orbitals, split into a lower $e_{g}^{\pi}$
doublet and an upper $a_{1g}$ singlet in the trigonal crystal field (CF) of
the corundum structure. In the low-$T$ monoclinic structure, the additional
low-symmetry CF component lifts the $e_{g}^{\pi}$ degeneracy.
At first glance, the experimental phase diagram shown in Fig.~\ref{fig:pd}
seems consistent with a positive (negative) chemical pressure exerted by Ti (Cr)
doping. It has therefore been routinely interpreted in terms of the standard
picture derived from the Hubbard model: upon applying positive (negative)
pressure, the band width $W$ increases (decreases), and so does the ratio
$W/U$, which controls the correlation strength and thus the tendency to form a
metal. This picture has been accepted in many previous works and appeared to
be supported~\cite{hel01,kel04,laa06,pot07} by a combination of density
functional theory (DFT) and dynamical mean-field theory (DMFT) without charge
self-consistency (i.e. one-shot scheme), whereupon the correlation-enhanced CF
splitting leads to a strong orbital polarization towards e$_{g}^{\pi}$ already
at stoichiometry. However, by finding still sizable $a_{1g}$ filling at zero
doping, this scenario has recently been questioned by angle-resolved
photoemission~\cite{vec16} and charge self-consistent DFT+DMFT
calculations~\cite{gri12,gri14,den14,leo15}. The chemical-pressure effect is
apparently too weak to explain key V$_{2}$O$_{3}$ phenomenology. In addition,
contrary to many statements in the literature, Ti doping leads to a volume
\emph{increase}~\cite{che82}, although less so than Cr.

Alternative causes for this metal-insulator transition (MIT) were
postulated~\cite{ric72,cas78,pfa02,tan02,bom04,fre06,men09,guo14}. Early on,
Rice and Brinkman~\cite{ric72} blamed doping-induced disorder and scattering
for the metallicity breakdown. Indeed, ion-irradiation experiments revealed
the sensitivity of the transition temperature to even a small concentration of
defects~\cite{ram15}. However, while such a scenario could explain why Cr
doping leads to an insulating state, it fails to explain the opposite effect
of Ti doping.

In this work, we show that the key effects of Cr or Ti doping cannot be
reduced to a chemical pressure. Instead, in a first approximation,
the Cr effect (reduced metallicity) is mostly structural (reduced 
local symmetry around Cr ion). The opposite effect of Ti is mostly electronic, 
an effective electron (and not hole!) doping of V$_{2}$O$_{3}$. This
conclusion is based upon a first-principles study employing state-of-the-art
charge-self-consistent DFT+DMFT calculations on relaxed supercells with Cr or
Ti doping. We emphasize that both a proper account of the structural and
charge self-consistency effects on the realistic defect chemistry, as well as
a precise inclusion of correlation effects, are essential for describing the
physics of V$_{2}$O$_{3}$.

\textit{Structural aspects.---} The starting point for the DFT+DMFT calculations
is the canonical two-formula unit-cell at stoichiometry (below called the
`stoichiometric' case), reported in Ref.~\cite{der70}. We first address the
\textit{global} structural effects of doping (chemical pressure) by
considering the minimal unit cell as reported by the experiment, $i.e.,$
averaged over the impurity disorder, for 2.8\% Cr~\cite{der70} and 3\%
Ti~\cite{che82} doping (`experimental-averaged' Cr or Ti, EA-Cr and EA-Ti,
respectively). For both dopings, the $c$-axis ($a$-axis) parameter is reduced
(increased) and the resulting volumes are larger than at stoichiometry.
However, the so-called \textquotedblleft umbrella\textquotedblright
distortion, $i.e.,$ the disparity of the two sets of O-metal-O angles,
changes only marginally in EA-Ti (from $15.8^\circ$ to $16.1^\circ$), 
but increases noticeably to $18.5^\circ$ with Cr.

Next, we construct 80-atom $2 \times 2 \times 2$ supercells with one of the 
32 V sites replaced by Cr or Ti, \textit{i.e.,} a 3.1\% doping, with the cell 
dimensions of the EA structures, and applied full symmetry-unrestricted atomic 
relaxation with DFT+U. Upon structural relaxation, the V ions near Cr shift 
toward the impurity, whereas the V-V distance of V ions farther away from the 
dopant increases, both in line with experiment~\cite{fre06,men09}. For Ti doping, 
on the other hand, V ions are repelled from the impurity. Second, local
monoclinic distortions are favored in Cr-doped V$_{2}$O$_{3}$, in agreement
with experiment~\cite{bom04,pfa02,men09}, but not in the Ti-doped compound.
Note that due to the limitations in manageable supercell sizes, the local
monoclinic distortions are described by a complete symmetry change of the
defect cell. Nevertheless, starting from the experimental corundum structure,
fixing the lattice parameters, and using a local-minimum relaxation algorithm
ensures that the resulting structure is a good approximation of the
true defect geometry. More details of the structural optimization are provided 
in the Supplemental Material~\cite{supp}.
\begin{figure}[t]
\centering
(a)\hspace*{-0.5cm}\includegraphics*[width=8.5cm]{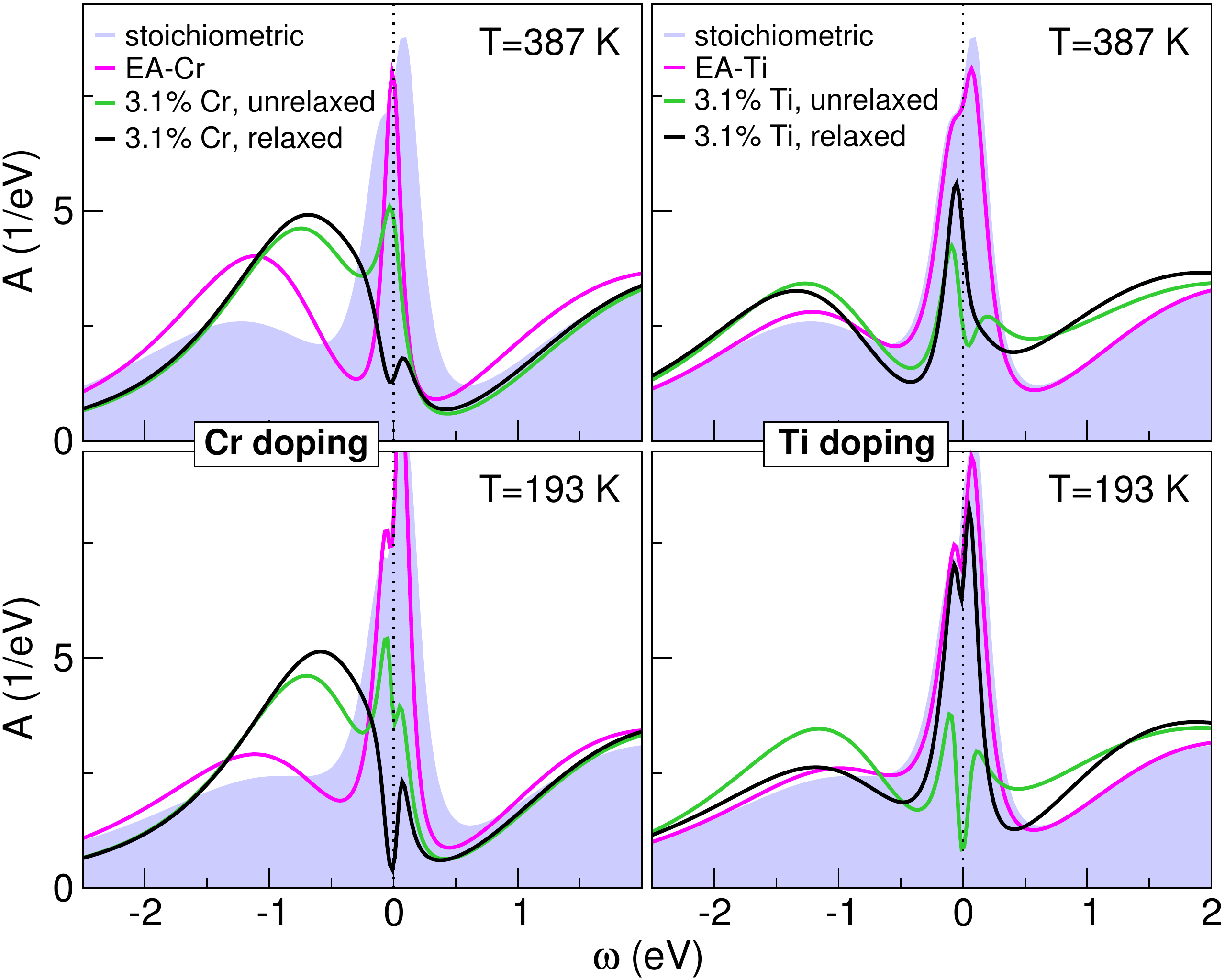}%
\\[0.2cm](b)\hspace*{-0.5cm}\includegraphics*[width=8.5cm]{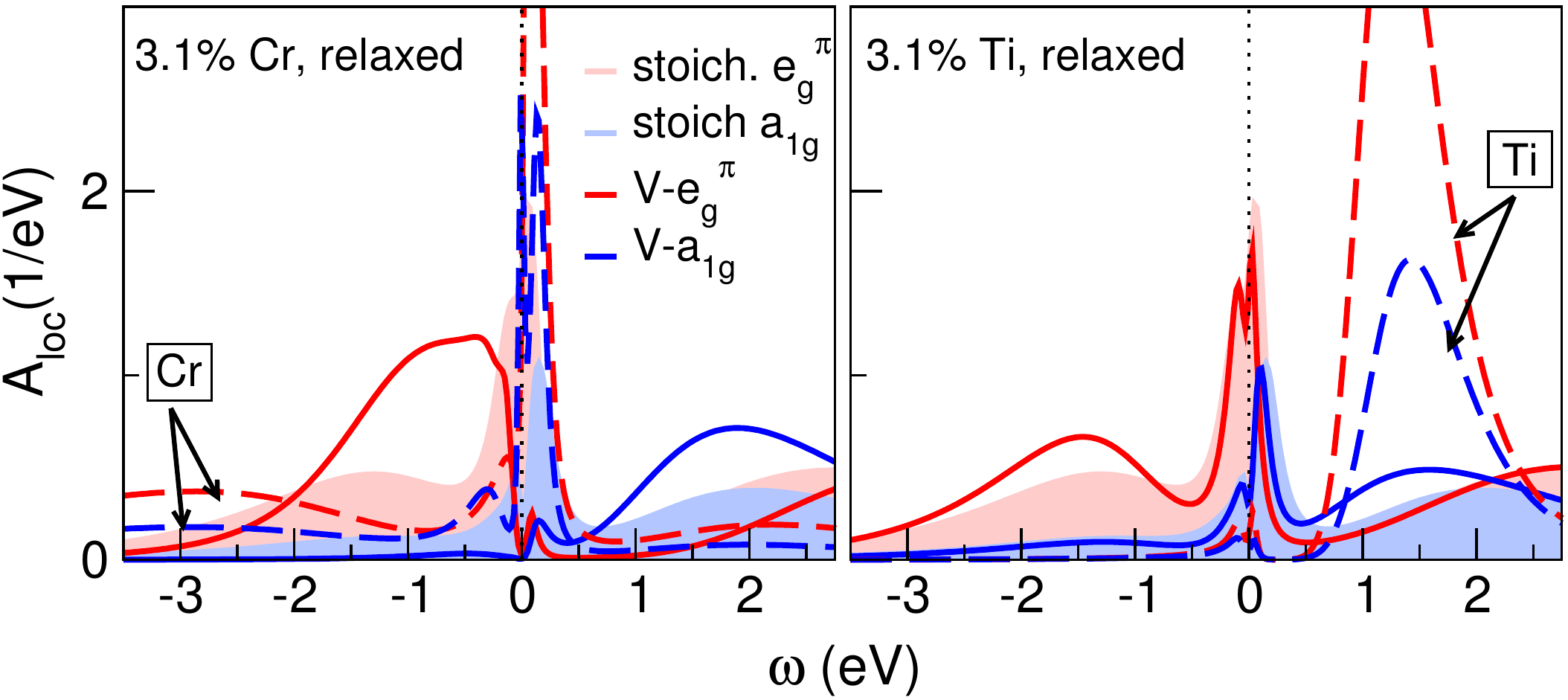}%
\caption{(Color online) DFT+DMFT spectral functions of stoichiometric and
doped V$_{2}$O$_{3}$. (a) Total spectral function at two $T$ for Cr doping
(left) and Ti doping (right). (b) Lattice-averaged V-$e_{g}^{\pi}$ and
V-$a_{1g}$ (full lines), and Cr and Ti $e_{g}^{\pi}$ and $a_{1g}$ (dashed
lines) local spectral function at $T=193\,$K, compared to orbital-resolved
stoichiometric $A_{\mathrm{loc}}$ (background).}%
\label{fig:dop-dmft}%
\end{figure}

A charge self-consistent DFT+DMFT~scheme~\cite{sav01,pou07,gri12} is applied
to thus obtained structures. The V-$t_{2g}$ triplet defines the correlated
subspace, with a Hubbard $U=5$\thinspace eV and Hund's-rule coupling 
$J_{\mathrm{H}}=0.7$\thinspace eV~\cite{supp}. Figure~\ref{fig:dop-dmft} shows 
the calculated spectral functions at $T=387$\thinspace K and $T=193$\thinspace K.
Stoichiometric V$_{2}$O$_{3}$ (Fig.~\ref{fig:dop-dmft}a shaded region)
displays a bare trigonal CF splitting of 162\thinspace meV. A moderate orbital
polarization towards $e_{g}^{\pi}$ is seen from the values of 
$n\equiv\{n_{e_{g}^{\pi}},n_{a_{1g}}\}=\{1.58,0.42\}$. It is a correlated metal 
with a renormalized quasiparticle (QP) peak and Hubbard bands, in line 
with photoemission observations~\cite{mo06,fuj11}. Based on the spectral weight 
of the lower Hubbbard band, the correlation strength appears to be somewhat larger 
at \textsl{higher} $T$. This is due to the small orbital-dependent
coherence-energy scale in V$_{2}$O$_{3}$~\cite{pot07,vec17,supp}; when the
temperature hits this scale, the low-energy QPs start to lose coherence and
spectral-weight transfer to the Hubbard sidebands occurs.

\begin{figure}[t]
\includegraphics[width=8.5cm]{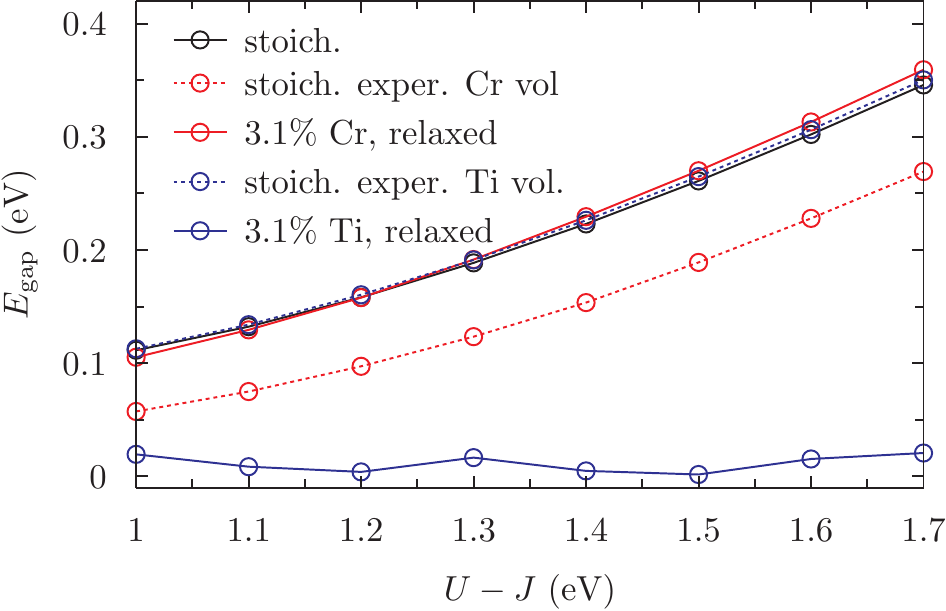}\caption{DFT+U band
gap as a function of $U_\mathrm{eff}$ for the pure stoichiometric system,
the pure system at volumes equivalent to the experimental 2.8\% Cr and 3\% Ti
doped volumes, and supercells with one atom of V substituted by Cr and Ti.}%
\label{fig:Eg_vs_U}%
\end{figure}

\textit{Ti doping.---} First, we investigate Ti doping with the various
unit-cells and supercells described above. For all cases, Ti-doped 
V$_{2}$O$_{3}$ remains metallic (Fig.~\ref{fig:dop-dmft}, right panels). 
We immediately observe that the Ti-$t_{2g}$ states (the dashed lines in 
Fig. ~\ref{fig:dop-dmft}b) are higher in energy than the corresponding V levels. 
So, formally, Ti$^{3+}$ loses its only 
electron and dopes the V states; (V$_{1-x}$Ti$_{x})_{2}$O$_{3}$ is thus 
electronically equivalent to V$_{2}$O$_{3}$ doped with $x/(1-x)$ electrons. At
low temperatures, as any doped Mott insulator, the system rapidly metallizes
(Fig.~\ref{fig:dop-dmft}a right panels). This is also 
 clearly seen in the DFT+U calculations (cf. Fig.~\ref{fig:Eg_vs_U}).
Ionically, the valence of the Ti impurity is hence ${4+}$, indeed in agreement
with experiment~\cite{mo-thesis}.

\textit{Cr doping.---} This case is more complex and Fig.~\ref{fig:dop-dmft}
shows our main results. Pure V$_{2}$O$_{3}$ set in the EA-Cr structure does
not exhibit a tendency to gap formation (pink curves in Fig.~\ref{fig:dop-dmft}a), 
contrary to earlier one-shot DFT+DMFT calculations~\cite{pot07}. The orbital 
polarization towards $e_{g}^{\pi}$ remains moderate (see the Supplemental 
Material for a complete account of the respective orbital 
occupations~\cite{supp}), unless unrealistically large lattice expansion is 
assumed~\cite{gri12,leo15}.
\begin{figure}[b]
\centering
\includegraphics*[width=8.5cm]{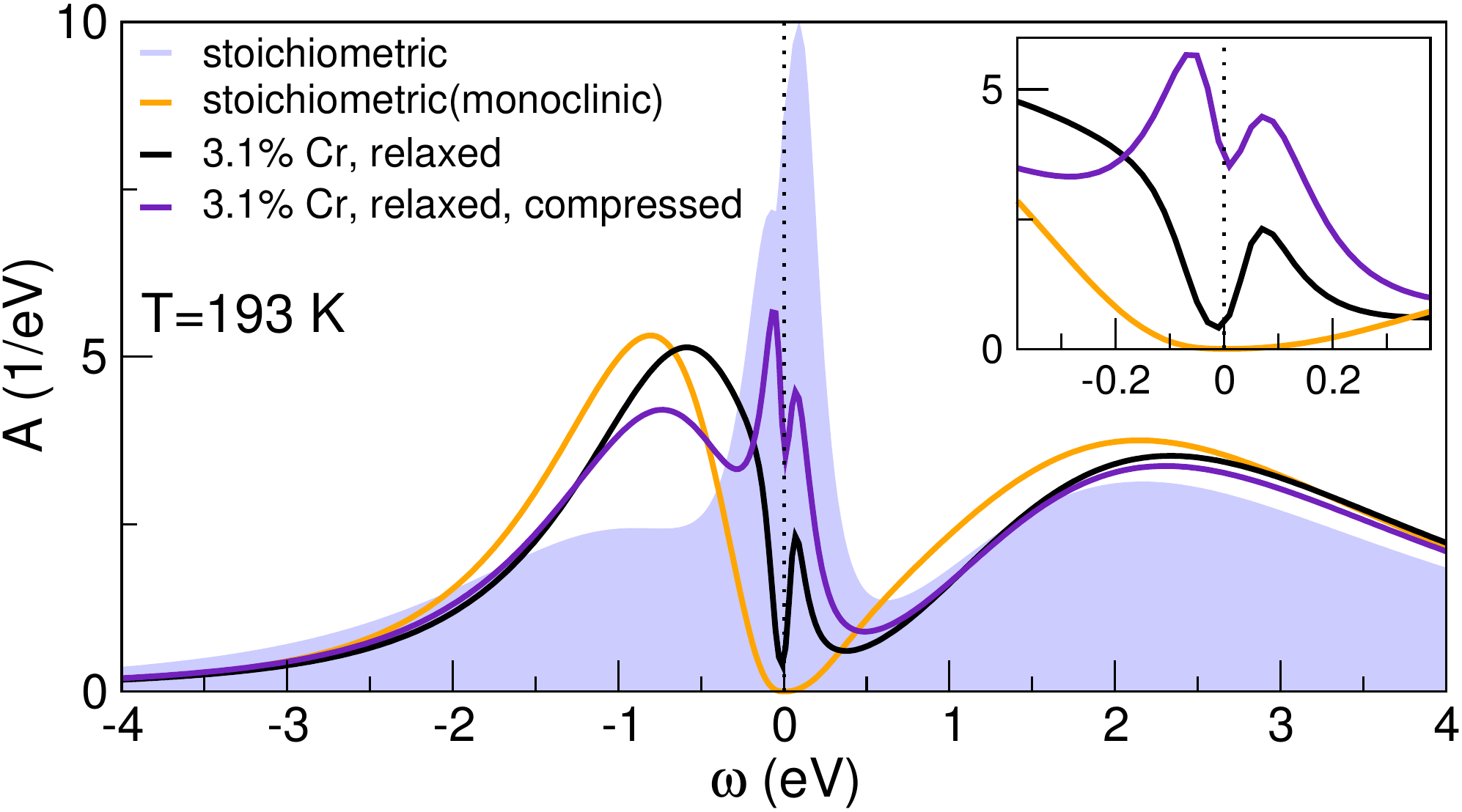}\caption{(Color online) DFT+DMFT
spectral functions at stoichiometry and with 3.1\% Cr doping. 'compressed'
Cr-doped case: relaxed supercell structure with stoichiometric lattice
parameters. 'stoichiometric(monoclinic)' case: experimental low-$T$ monoclinic
structure without doping, here at $T=193$\thinspace K treated as paramagnetic,
with a monoclinic bare CF splitting $\Delta_{e_{g}^{\pi}}=29$\thinspace meV
and an orbital filling $n=\{1.96,0.04\}$. Inset: low-energy blow up.}%
\label{fig:mono}%
\end{figure}
However, a complete structural relaxation with impurity-induced local
monoclinic distortions leads to a gap opening in the paramagnetic phase as
shown by the disappearance of spectral weight at low energy (\textit{3.1\% Cr,
relaxed}, black curves). The gap size of $\sim100\,$meV (at $T=193$\thinspace
K) is in excellent agreement with experiment~\cite{bar70,mo06}. While even the
simple replacement of a single V ion by Cr already leads to a reduction of
low-energy spectral weight and strengthening of correlations~\cite{gri14} 
(\textit{3.1\% Cr, unrelaxed}, green curves), structural relaxation is required for 
opening the gap. From analyzing the local spectral functions we find a complete 
MIT, \textit{i.e.} no sign of orbital-selective behavior~\cite{laa06,leo15} as
shown in Fig.~\ref{fig:dop-dmft}b, where the lattice-averaged V-$e_{g}^{\pi}$
and V-$a_{1g}$ function for the \textit{3.1\% Cr, relaxed} structure (solid
line) is plotted. Figure~\ref{fig:mono} shows that under external pressure 
the PI phase recovers metallicity, in agreement with experimental work by
Rodolakis~\textsl{et al.}~\cite{rod10}. Importantly, this stabilized metallic 
phase is not identical to the original stoichiometric one. Instead, it shows a 
sizable pseudogap and a large part of the Cr-induced $e_{g}^{\pi}$ polarization 
is preserved. For comparison, the experimental low-$T$ monoclinic structure at
stoichiometry is easily Mott insulating already in paramagnetic DFT+DMFT
(see Fig.~\ref{fig:mono}, orange line), with a sizable charge gap of about 0.4\,eV.

Obviously, Cr behaves completely differently with respect to the V matrix
of V$_2$O$_3$
compared to Ti. This can be  traced to several effects. First, the
one-electron Cr-$t_{2g}$ level is \textit{below } the entire V-$t_{2g}$ 
manifold (as shown from DFT+U calculations in the Supplemental 
Material~\cite{supp}), and reflected in a broad impurity
state centred at $-2.8$\,eV in Fig.~\ref{fig:dop-dmft}b (which is reminiscent
of a similar feature around $-2$\,eV in photoemission~\cite{mo06}).
Formally, the Cr$^{3+}$ ion has three electrons, which constitute a spin $S=3/2$ 
state below the V-$t_{2g}$ states. The formal occupancy of the V-$e_{g}^{\pi}$ 
states remains essentially unaltered, so one has to look at more subtle, 
structural effects. First, the stronger and more asymmetric trigonal distortion 
(\textquotedblleft umbrella distortion\textquotedblright, UD) of the EA-Cr 
structure, compared to stoichiometric V$_{2}$O$_{3},$ creates a 
stronger $e_{g}^{\pi}-a_{1g}$ splitting. With the Cr impurity but
without any further structural changes, the orbital polarization towards 
$e_{g}^{\pi}$ is then already very strong with $n=\{1.87,0.10\}$. 
Yet only the additional 
monoclinic CF triggered by the explicit local-symmetry breaking effects of Cr 
doping finally drives the MIT. Note that this additional CF splitting is only 
$\Delta_{e_{g}^{\pi}}\sim5$$-$$20$\thinspace meV, depending on the V site, but 
it proves relevant to establishing the PI phase. Multi-orbital Mott transitions 
driven by subtle CF splitting have been studied in various model-Hamiltonian
schemes~\cite{man02,kit11}. 
\begin{figure}[t]
\includegraphics*[width=8.5cm]{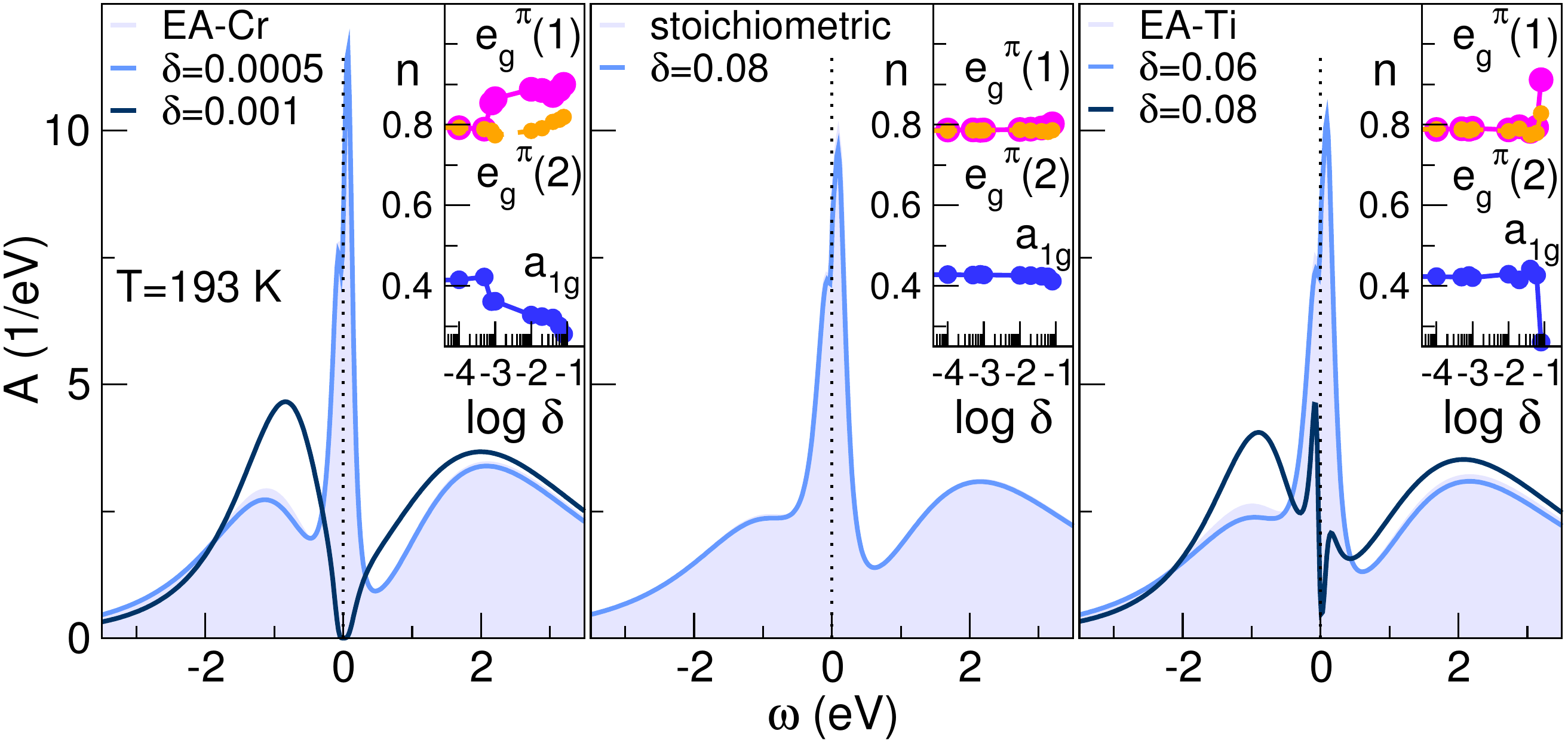}\caption{(Color online) DFT+DMFT data
for the small-unit-cells EA-Cr (left), stoichiometric (middle) and EA-Ti (right) 
with symmetry-breaking charge-density modifications: selected total 
spectral functions and orbital occupations (inset). The quantity $\delta$ marks the
O-nuclei charge deviation on site $i$, i.e. $Z_{i}({\mathrm{\tilde{O}}}%
)=Z_{i}({\mathrm{O}})+\delta_{i}$ for distributed
$\delta_{i}=(-1)^{i}\,\delta$ with $\sum_{\mathrm{cell}}\delta_{i}%
=0$~\cite{supp}.}%
\label{fig:delta}%
\end{figure}

However, importantly, the Cr-induced UD plays its own, quite essential role, 
albeit indirectly. In order to 
address the effect of the structure symmetry alone, we have performed calculations 
using the high-symmetry EA-Cr, stoichiometric and EA-Ti structures 
(which differ mostly by the degree of the UD), and artificially lowered the 
symmetry by randomly replacing the ligand oxygens with pseudo-oxygen of
nuclear (and electron) charge $8\pm\delta$ (see the Supplemental Material
for more details~\cite{supp}). The result of this experiment
is shown in Fig.~\ref{fig:delta}. While EA-Cr transforms into a Mott insulator
at an already tiny symmetry-lowering field, the other metallic structures remain 
comparatively robust to such a field. Thus, the same high-symmetry corundum 
structure is much more sensitive to symmetry lowering if it has a stronger UD 
asymmetry, as is the case in EA-Cr.

This effect is the leading source of the insulating behavior in Cr-doped
samples, but is by far not the only effect present. For instance, charge 
fluctuations and Kondo-like screening of the additional Cr spin generate an 
empty QP peak (dashed curve in Fig.~\ref{fig:dop-dmft}b, left panel) just at 
the bottom of the upper gap edge, which can, in principle, be visible in 
inverse photoemission. Notably, this feature should lead to a {\sl decrease} of
the charge gap at low temperatures with Cr doping. And indeed, this originally
counterintuitive gap development is observed in photoemission~\cite{mo06}.

\textit{DFT+U calculations.---} We have established above that the charge
self-consistent DFT+DMFT supercell calculations, which include local
correlation effects, are capable of reproducing the experimental trends for
both kinds of doping (Cr and Ti) when distortions around impurities and charge
alteration are accounted for. At this point, it is instructive to ask which
effects introduced by doping are already captured by an effective one-electron
approach. To this end, we analyze here whether the static mean-field DFT+U
method shows the right trends of increasing (reducing) the low-$T$ gap when
doping V$_{2}$O$_{3}$ by Cr (Ti). We use DFT+U with $U_\mathrm{eff}=U-J=1.5$~eV to
relax (with respect to atomic position and cell shape, but not volume) our
minimal two-formula unit cells at the experimental Cr and Ti doped volumes.
Additionally, our already relaxed 80-atom
supercells with impurities are further optimized with that protocol in order
to realize now the low-$T$ scenario. For all these structures, the band gap as
a function of $U_\mathrm{eff}$ is computed (Fig.~\ref{fig:Eg_vs_U}). When optimized
at the volume of the experimental Ti-doped material, the undoped system has
essentially the same gap as at the pure material volume, while at the
experimental Cr-doped volume the gap goes \textit{down} by about
0.07~eV. However, the same procedure applied to the relaxed
supercells of the doped systems leads to a completely different behavior: for
Ti doping the gap is reduced to zero (within computational noise), while for
Cr doping the gap increases slightly over the original pure material's band
gap. We should point out that, for the purpose of uniformity, we have used the
same optimization procedure for all compound scenarios. While a much larger
$U_\mathrm{eff}$ is needed to open a gap in the high-$T$ corundum structure, in the
optimized low-symmetry $T=0$ AF structure the gap opens already at
$U-J\lesssim1$~eV. These results demonstrate that the opposite, and
qualitatively correct, trends of Cr vs.\ Ti doping are captured at the
effective one-electron level. However, correlation effects as implemented in
DMFT are essential to describe the PM-PI transition at finite temperatures and
distinguish between the less correlated undoped behavior and the more
correlated Cr-doped case.

\textit{Conclusions}.--- Our charge self-consistent results prove that the
intriguing Cr- and Ti-doping phase diagram of V$_{2}$O$_{3}$ is not driven by
chemical pressure. In fact,  contrary to the common misconception, the
principal factors controlling the phase diagram are qualitatively different 
for the two impurities. On the Cr side, the main factor is a local structural
distortion affecting the crystal field, while on the Ti side it is effective
e-doping by virtue of Ti acquiring a valence of 4$+$. Increasing the
``umbrella'' distortion in the average
corundum structure by Cr doping is not $per$ $se$ sufficient to trigger a
paramagnetic insulator, but it makes the system more sensitive to further
symmetry lowering on the local level. In the simplest approximation, Ti produces a
charge doping and Cr does not, while Cr produces a structural distortion
conducive to insulating behavior and Ti does not. 

Our findings elucidate the major factors controlling the phase diagram of
vanadium sesquioxide, and should also be relevant to interpret recent 
V$_{2}$O$_{3}$ thin-film studies~\cite{maj17,tho18}. We furthermore observe 
that whereas these effects are importantly enhanced by electron correlations, 
the described trends appear already at the one-electron level within the static 
mean-field DFT+U framework.

\begin{acknowledgments}
We thank J. W. Allen, J. D. Denlinger and I. K. Schuller for helpful discussions. 
FL acknowledges financial support from the DFG project LE 2446/4-1. IIM and NB
were supported by ONR through the NRL basic research program. RV was supported
by the DFG through SFB/TR 49. DFT+DMFT computations were performed at the
JURECA Cluster of the J\"{u}lich Supercomputing Centre (JSC) under project
number hhh08.
\end{acknowledgments}

\bibliographystyle{apsrev}
\bibliography{bibextra}

\clearpage

\begin{center}
{\Large Supplemental Material}
\end{center}

\section{Electronic structure methods}

\textit{DFT+U}.--- We used the Vienna ab-initio software package
(VASP)~[1] version 5.4.1 within the generalized-gradient
approximation given by the Perdew-Burke-Enrzerhof (PBE) exchange-correlation
functional~[2], and the regular VASP projector-augmented waves (PAW)
with 6 electrons in the valence for O, and the $p$-valence PAWs for V, Cr, and
Ti, with 11, 12, and 10 electrons in the valence, respectively. All
calculations used a plane-wave energy cutoff of 400\thinspace eV, and a
k-point mesh equivalent to $4\times4\times4$ for the 10-atom primitive
two-formula unit cell. A spherically-averaged Hubbard correction with the
fully-localized limit double-counting subtraction~[3] is used. The
value $U-J=1.5$\thinspace eV is utilized for the structural relaxation. While
smaller than this value of $U-J$ is needed to reproduce the band gap (see
discussion in the main text), we found it to yield a good equilibrium
structure compared to experiment for the undoped compound.

\textit{DFT+DMFT}.--- The charge self-consistent combination of density
functional theory and dynamical mean-field theory is used. For the DFT part, a
mixed-basis pseudopotential method~[4,5], built on
norm-conserving pseudopotentials as well as a combined basis of localized
functions and plane waves is utilized. The PBE functional is employed. Within
the mixed basis, localized functions for transition-metal $3d$ states as well
as for O($2s$) and O($2p$) are used in order to reduce the energy cutoff
$E_{\mathrm{cut}}$ for the plane waves. The correlated subspace consists of
the effective transition-metal $t_{2g}$ Wannier-like functions $w_{n}(t_{2g}%
)$, i.e. is locally threefold. The $w(t_{2g})$ functions are obtained from the
projected-local-orbital formalism~[6,7], using as projection
functions the linear combinations of atomic $t_{2g}$ orbitals, diagonalizing
the transition-metal $w_{n}(t_{2g})$-orbital-density matrix. The coupled
single-site DMFT impurity problems in the V$_{2}$O$_{3}$ crystal cells are
solved by the continuous-time quantum Monte Carlo scheme~[8,9] as
implemented in the TRIQS package~[10,11]. A double-counting
correction of fully-localized type~[3] is utilized. To obtain the
spectral information, analytical continuation from Matsubara space via the
maximum-entropy method is performed. About 50-100 DFT+DMFT iterations (of
alternating Kohn-Sham and DMFT impurity steps) are necessary for full
convergence of the supercell problems.

Former DFT+DMFT studies described the paramagnetic Mott transition in Cr-doped
V$_{2}$O$_{3}$ utilizing the experimental-averaged (EA) case and mostly employing 
two sets of local Coulomb parameters. Either $U=4.2\,$\,eV, $J_{\mathrm{H}}=0.7\,$eV 
were used~[12] or $U=5.0\,$eV, $J_{\mathrm{H}}=0.93\,$\,eV~[13,14,15]. 
While we confirm the MIT via the
correlation-enhanced CF splitting in one-shot DFT+DMFT for the former, both
sets appear to result in too weak correlation strength in the more elaborate
charge self-consistent framework. This is understandable, since charge
self-consistency enables reduced correlations due to charge redistribution
also incorporating ligand states. We hence choose $U=5.0\,$eV, $J_{\mathrm{H}%
}=0.7\,$eV to promote the correlation strength to experimentally-known
measures. A $3\times3\times3$ k-point mesh is employed for the supercell calculations.

\section{Structural approach}
\indent \textit{Basic cells}.--- For the conventional primitive unit cells of
V$_{2}$O$_{3}$ we started off from the various experimental data (see
Tab.~\ref{tab:explat} for a summary).

\begin{figure}[t]
\centering
\includegraphics*[height=8cm]{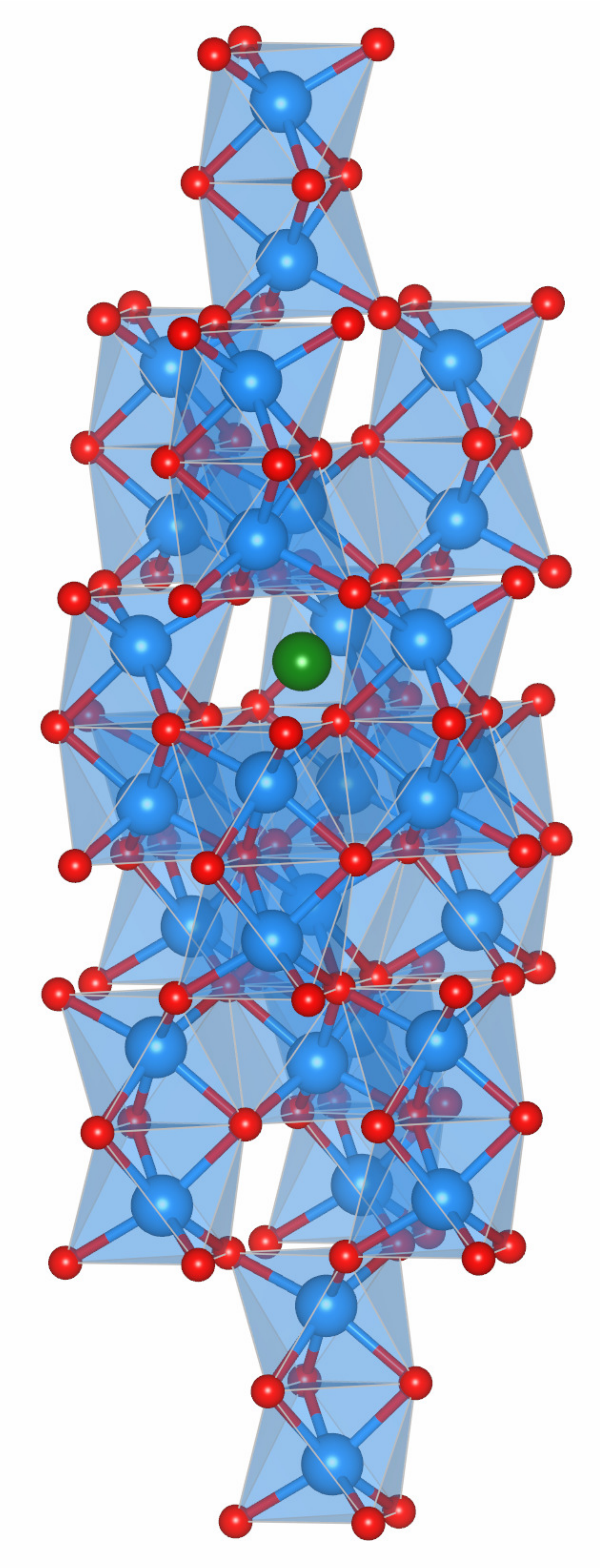} \caption{(Color online) 80-atom
supercell to model paramagnetic V$_{2}$O$_{3}$ with 3.1\% dopants. Ti: blue,
O: red, dopant: green. The relaxed nearest-neighbor distances between dopant
and V amounts to 2.715\thinspace\AA \thinspace\ for Cr doping and to
2.789\thinspace\AA \thinspace\ for Ti doping.}%
\label{fig:structure}%
\end{figure}

\textit{Supercells}.--- We have utilized supercells obtained by doubling the
corundum unit cells in each crystallographic directions. Thus obtained
supercells correspond to the net formula V$_{32}$O$_{48}$. Substituting one V
by Ti or Cr results in 3.125\% doping, comparable with the experimentally
interesting ranges (see Fig.~\ref{fig:structure}). For each case (Cr- or
Ti-doped) we used experimentally-determined cell lattice parameters and
volumes. In further structural relaxations within DFT+U for generating
configurations for DFT+DMFT, the atomic coordinates are optimized with
$U_{eff}=1.5$\thinspace eV. Introducing a single dopant atom into the 80-atom
corundum supercell leads to 16 symmetry-inequivalent DMFT impurity problems
associated with the transition-metal sites.
\begin{table}[b]
\begin{ruledtabular}
\begin{tabular}{l|cccc}
structure           & $a,b$  & $c$  & V (\AA$^3$) & NN-V \\ \hline
stoich. corundum    & 4.952       & 14.003 &  99.107  & 2.697 \\[-0.1cm]
($T=290\,$K), Ref.~[16] &  &  & & \\
stoich. monoclinic  & 7.255  &  5.548 &  99.969  & 2.745  \\[-0.1cm]
($T=148$\,K), Ref.~[17] & 5.000 &  & & \\
2.8\% Cr-doped corundum   & 4.999       & 13.912 & 100.341  & 2.746 \\[-0.1cm]
($T=290\,$K), Ref.~[16] &  &  & & \\
3\% Ti-doped corundum   & 4.963       & 14.000 &  99.558  & 2.706  \\[-0.1cm]
($T=290\,$K), Ref.~[18] &  &  & & \\
\end{tabular}
\end{ruledtabular}
\caption{Utilized experimental V$_{2}$O$_{3}$ crystal lattice data (in \AA ).
The distance 'NN-V'' corresponds to the nearest-neighbor V (dimer) distance
along the $c$-axis. The correspondance with the main-text nomenclature is as
follows: 'stoich. corundum';'stoichiometric', 'stoich. monoclinic';'low-$T$
monoclinic', 'Cr-doped corundum';'EA-Cr' and 'Ti-doped corundum';'EA-Ti'.}%
\label{tab:explat}%
\end{table}

\textit{Describing local monoclinic distortions}.--- Global monoclinic
distortions occur in the insulating low-$T$ regime of V$_{2}$O$_{3}$ and are
here representatively treated for the stoichiometric case. There,
experimentally a four-formula unit-cell with $I2/a$ space group is
realized~[17]. On the other hand, true local monoclinic distortions,
especially at finite $T$, are hard to describe in a well-defined manner within
conventional first-principles approaches dealing with feasible supercell
sizes. However together with the corresponding electronic-structure treatment,
our relaxed supercells are already good structural approximants to V$_{2}%
$O$_{3}$ phases with possible local monoclinic distortions, due to the three
following reasons.
First, the corundum structure serves as starting point for the structural
relaxation. Standard relaxation of the atomic positions will most likely
converge to the nearest minimum in the complex total-energy landscape. From a
numerical viewpoint, there is no obvious reason to believe that such an
unbiased relaxation will directly lead to the true structral ground state at
low temperature. Instead, a local energy minimum will be realized, that is
energetically comparable to true local-relaxation modes describable only in
supercells of size $\sim10^{3}$ and more atoms. Second, the supercell lattice
parameters are fixed to their experimental values (for 'EA-dopant') at
\textsl{elevated} temperature. Hence, the very region where local-monoclinic
relaxation may occur is imprinted in the supercell structures.
\begin{figure}[b]
\centering
\includegraphics*[width=7.5cm]{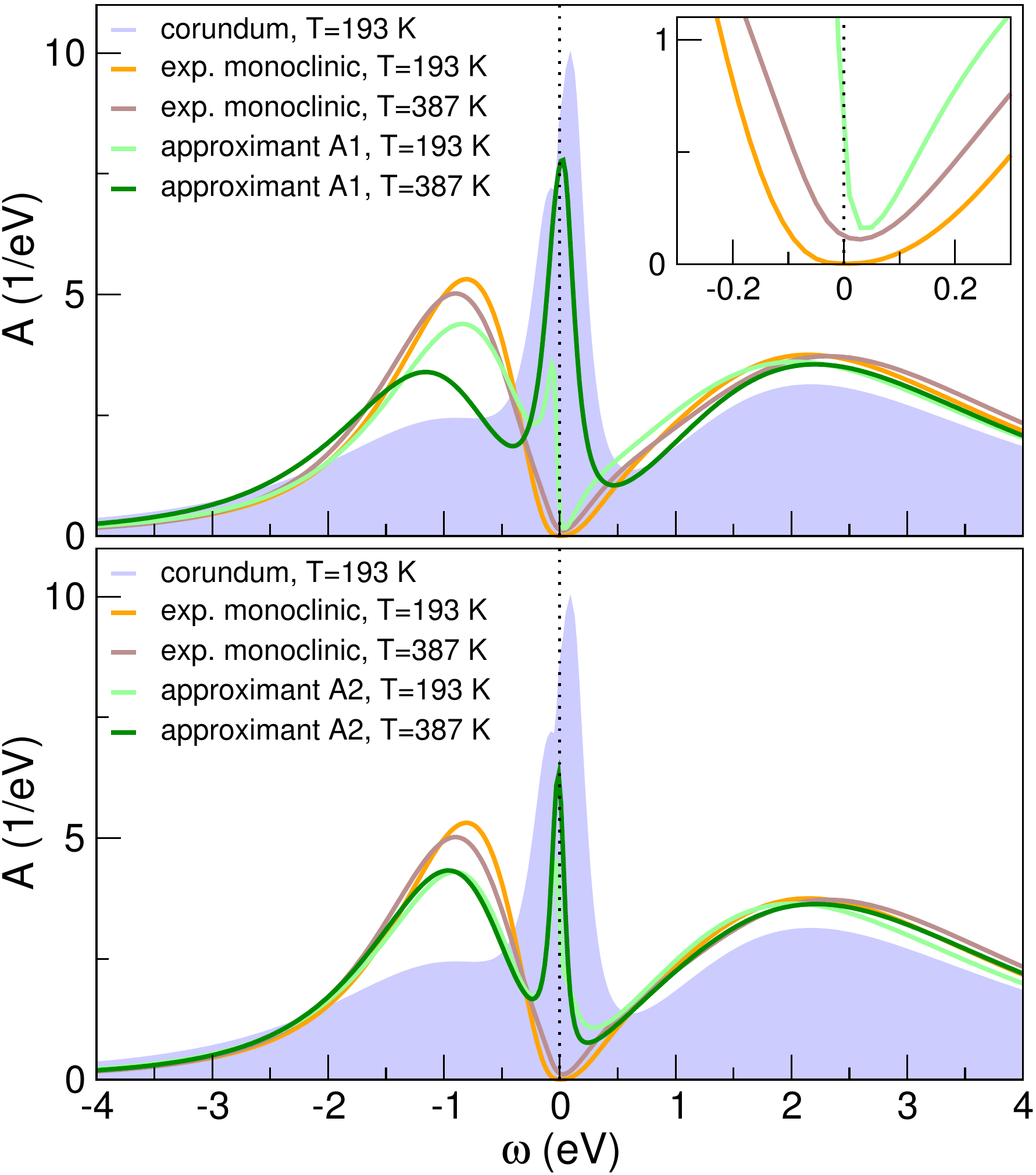}\caption{(Color online) Correlated
spectral functions of the different monoclinic approximants A1 (top) and (A2)
bottom, see text, compared to the ones of experimental monoclinic 
V$_{2}$O$_{3}$ at stoichiometry.}%
\label{fig:coru-mono}%
\end{figure}
Third, in the supercell DFT+DMFT calculations, the self-energy symmetry among
the V sites is tied to the unrelaxed dopant corundum structure. In other
words, further possible monoclinic symmetry breaking is computationally not
accompanied with an increase of symmetry-inequivalent V self-energies. This is
to (partly) restore the global corundum symmetry at high temperature. Formally
inequivalent V self-energies, resulting from monoclinic symmetry breaking, are
thus averaged in the supercell calculations.

To underline our reasoning, Fig.~\ref{fig:coru-mono} displays the DFT+DMFT
result of applying the first and second argument to stoichiometric V$_{2}%
$O$_{3}$. There, the low-$T$ structure is experimentally known and its
spectral function can be compared to the ones of structural approximants
obtained from relaxation of the known stoichiometric structure at higher $T$.
The monoclinic low-$T$ structure is verified to be Mott-insulating even from
above the experimental $T_{\mathrm{N}}$ up to the crossover temperature.
Clearly, structural changes are necessary to drive a metal-insulator
transition. Two structural approximants are constructed, starting from the
stoichiometric high-$T$ corundum structure and fixing the lattice parameters.
The first approximant (A1) results from an unbiased straightforward
optimization of the atomic positions as well as the cell shape. The second
approximant (A2) is obtained from optimizing atomic positions and cellshape of
a doubled corundum unit-cell and additionally fixing inversion symmetry (which
old also in the true $I2/a$ structure). Figure~\ref{fig:coru-mono} shows that
none of the approximants reaches the gap size of the experimental structure.
While A1 displays a much smaller gap at low $T$, A2 remains metallic even at
$T=193$\,K. This proves that structural relaxation from the corundum high-$T$
structure does not lead to the true low-$T$ monoclinic structure, but to some
effective structure that should be a reasonable approximant to mimic
local-monoclinic distortions in the corundum phase. We note that the obtained
monoclinic CF values (i.e. energy splitting between the $e_{g}^{\pi}$ states)
for approxmiants A1 ($\Delta_{e_{g}^{\pi}}=11$\,meV) and A2 ($\Delta
_{e_{g}^{\pi}}=6$\,meV) are much smaller than for the experimental structure
($\Delta_{e_{g}^{\pi}}=29$\,meV).

\section{DFT electronic structure of Chromium-doped V$_{2}$O$_{3}$}
\begin{figure}[b]
\includegraphics*[width=8.5cm]{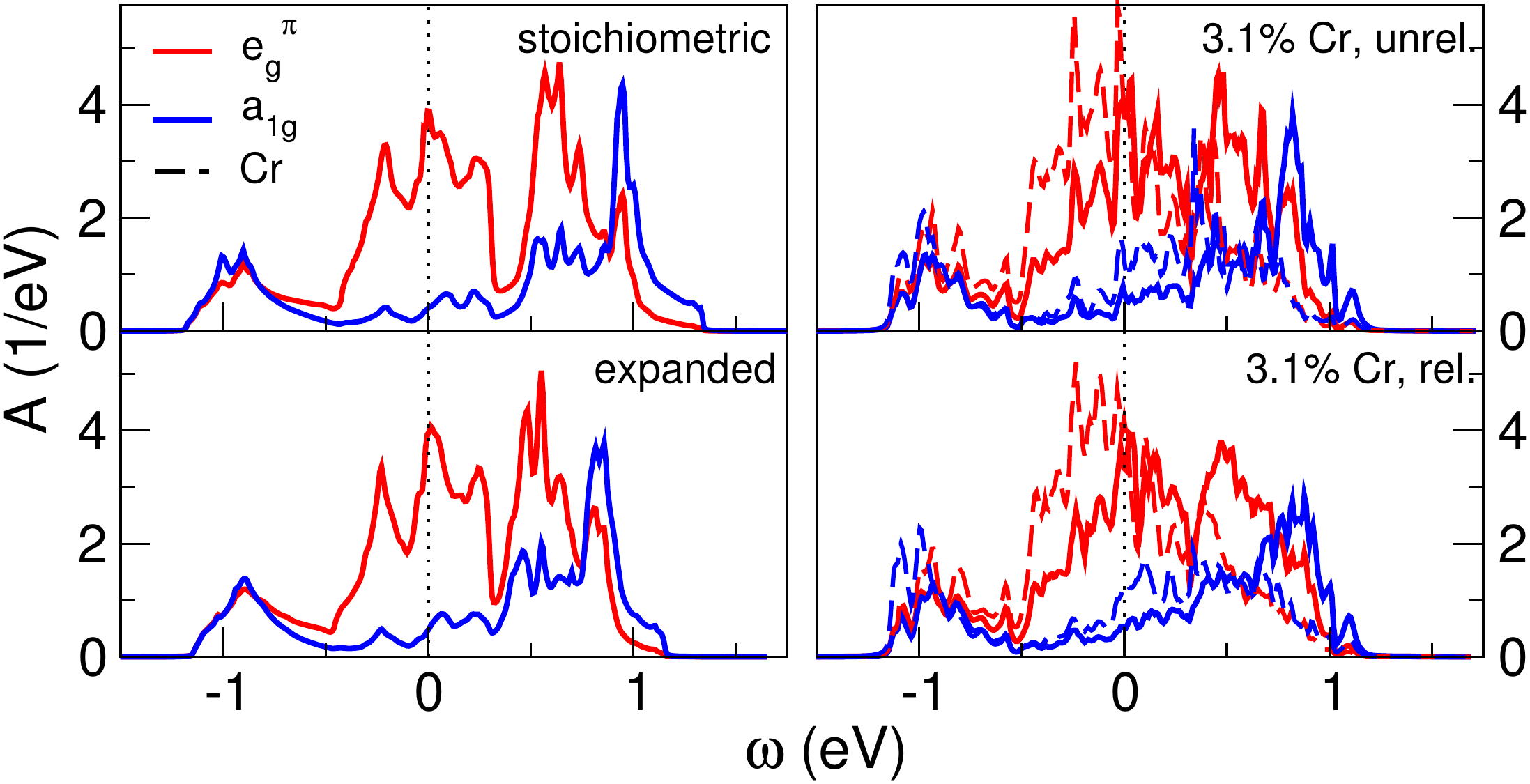}\caption{(Color online) DFT
account of stoichiometric and Cr-doped V$_{2}$O$_{3}$ via lattice-averaged V
and Cr $e_{g}^{\pi}$ and $a_{1g}$ lDOS (normalized to one site).}%
\label{fig:crdop-dft}%
\vspace*{0.4cm}
\includegraphics*[width=4.15cm]{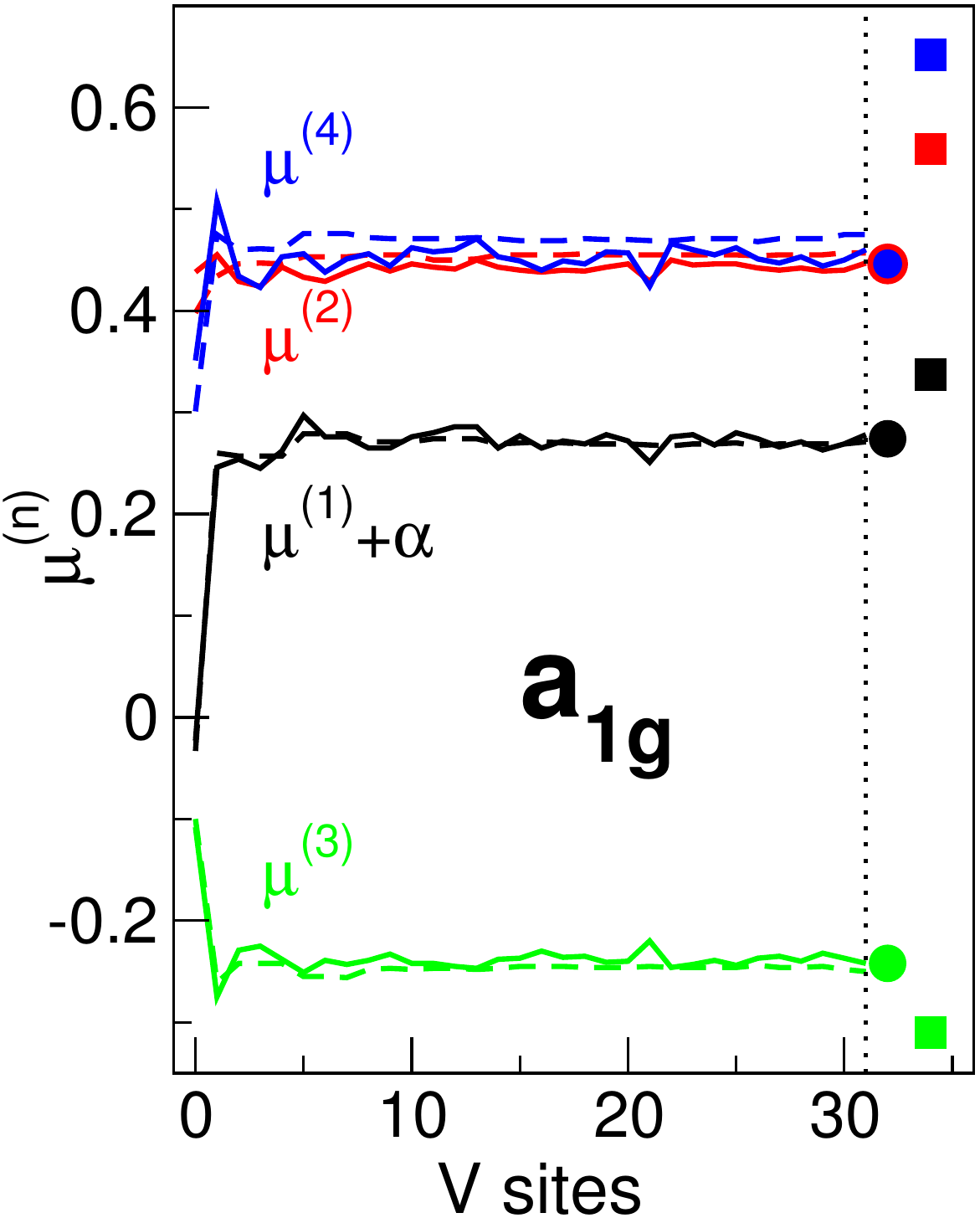}\hspace*{0.1cm}
\includegraphics*[width=4.15cm]{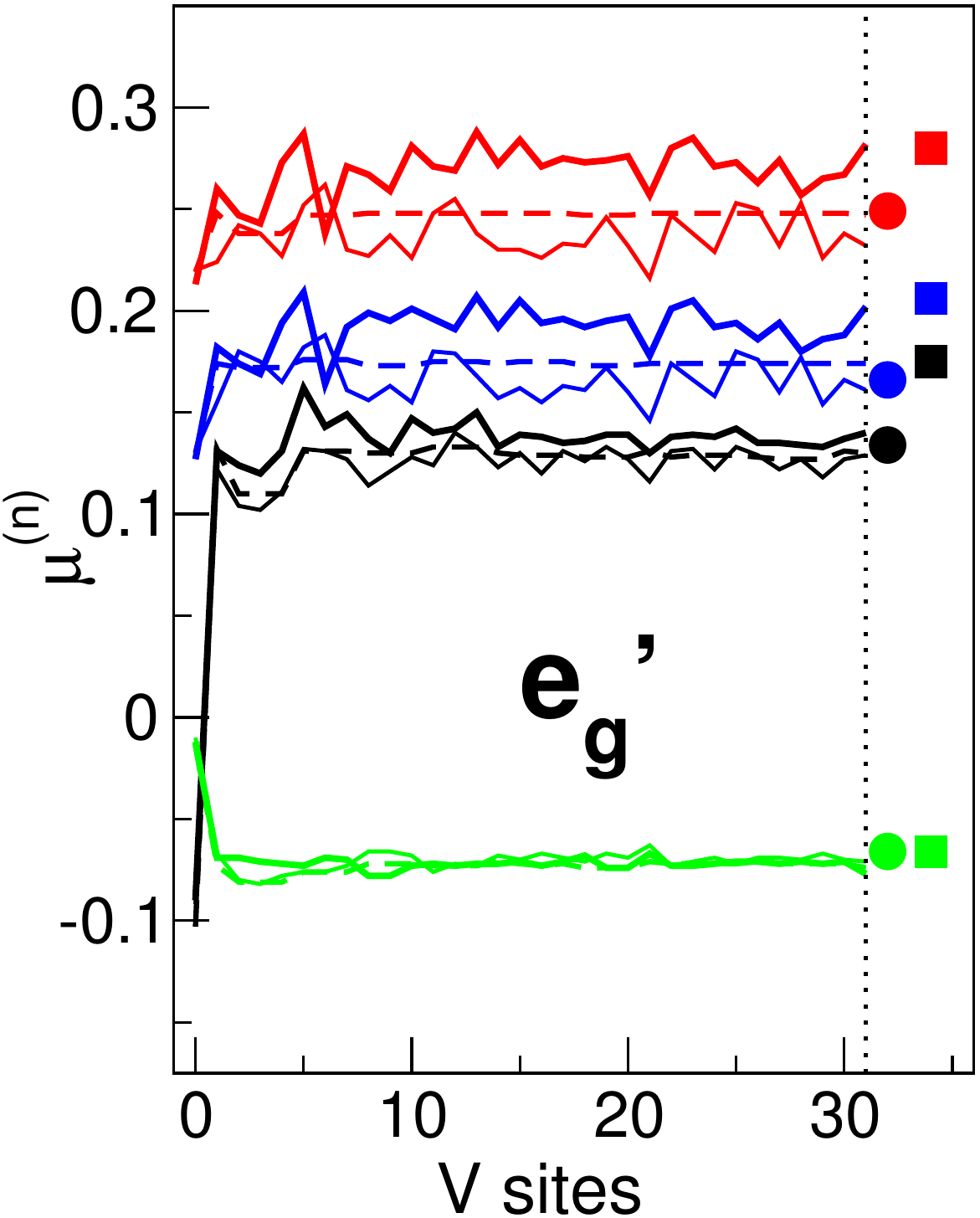}\caption{(Color online) Moment
analysis for $a_{1g}$ and $e_{g}^{\pi}$ character with respect to V sites
distant to Cr impurity. Solid(Dashed) lines: relaxed(unrelaxed) Cr-doped
supercell. Squares(Circles): stoichiometric(EA) V$_{2}$O$_{3}$. Since V
sites in the stoichiometric and EA cases are equivalent, respectively,
only a single moment is associated with each case.}%
\label{fig:moments-dft}%
\end{figure}
\textit{Orbital-resolved density of states}.--- On the DFT level, the
orbital-resolved electronic structure of Cr-doped vanadium sesquioxide is as
follows. Roughly speaking, the $a_{1g}$ orbital mainly points along the
crystallographic $c$-axis, where the corundum structure designates pairs of V
atoms (though not readily in a dimer-like configuration as e.g. in VO$_{2}$~[19]. 
Both $e_{g}^{\pi}$ orbitals are more extended in the
honeycomb $ab$-plane. The local density of states (lDOS) of low-energy
$(e_{g}^{\pi},a_{1g})$ kind for the different structural cases is displayed in
Fig.~\ref{fig:crdop-dft}.
At stoichiometry with bandwidth of order 2.5\thinspace eV, the $a_{1g}$ lDOS
displays strong bimodal character, derived from bonding-antibonding-like V-V
chemistry. The $e_{g}^{\pi}$ states are more highly filled and have sizable
weight close to the Fermi level $\varepsilon_{\mathrm{F}}$. Subsequent
modifications due to lattice expansion and Cr doping seem minor on this
integrated DFT level. The bandwidth shrinks to $\sim2.3$\thinspace eV in the
EA case and the $e_{g}^{\pi}$ peak structure at $\varepsilon
_{\mathrm{F}}$ becomes less split with Cr. The Cr-$e_{g}^{\pi}$ defect states
have more weight in the occupied part due to the larger Cr valency and a
closer look reveals additional peaks at the lower $a_{1g}$ band edge.

\textit{Moment analysis}.--- In order to obtain a most-compact insight on the
defect influence from the DFT level, we performed a site- and orbital-resolved
analysis of the electronic structure moments $\mu^{(n)}$. Here, the moments
theorem~[20,21], making use of the lDOS $D_{im}$ for
transition-metal site $i$ and orbital $m=(e_{g}^{\pi},a_{1g})$, takes on the
form
\begin{equation}
\mu_{im}^{(n)}=\int_{\mathrm{band}}d\varepsilon\;(\varepsilon-\alpha_{m}%
)^{n}\,D_{im}(\varepsilon)\quad,
\end{equation}
whereby $\alpha_{m}:=\langle im|H^{\mathrm{(KS)}}|im\rangle$ denotes the
onsite matrix element of the Kohn-Sham Hamiltonian. The $n$th moment is the
sum of all electron paths with $n$ hops on the lattice starting and ending at
site $i$. Therefore, inspection of the various moments $\mu_{im}^{(n)}$
enables a well-defined account of the delicate electronic and structural
defect-mediated modifications. Figure~\ref{fig:moments-dft} shows the
low-energy moments up to $n=4$ with respect to the V sites in distance to the
Cr impurity. The zeroth moment is unity, while the first moment vanishes due
to the cancellation by $\alpha_{m}$. We thus plot the crystal-field levels
$\mu^{(1)}+\alpha_{m}$, revealing the largely different values at the Cr site
and the slightly increased vanadium values further away upon structural
relaxation compared to the EA structure. Yet the averaged CF splitting
in the defect cases differs only marginally from the EA case
$\Delta_{\mathrm{e}}\sim140\,$eV, both being somewhat smaller than at
stoichiometry $\Delta_{\mathrm{s}}\sim160\,$eV. Close to Cr, there is a
sizable variation of the CF $e_{g}^{\pi}$ levels, which then split into two
branches with distance to the impurity. Thus importantly, the impurity with
subsequent structural relaxation leads to a global lifting of the
V-$e_{g}^{\pi}$ degeneracy due to symmetry breaking. There is partly also some
$e_{g}^{\pi}$ splitting in the unrelaxed case, but much smaller (not shown).
The second moment $\mu_{im}^{(2)}$ is directly related to the bond energy
stemming from site $i$ and orbital $m$. It becomes small close to Cr on the
fixed corundum lattice, but is reinforced upon relaxations. Also here, the
$e_{g}^{\pi}$ sector splits in values above and below the EA case
further away from Cr. The skewness is associated with the third moment and the
bimodal character with the fourth one. The latter displays a sizable peak in
the $a_{1g}$ channel for the V pairing with Cr, signaling an enhanced
bonding-antibonding character. Concerning relevant length scales, the
$e_{g}^{\pi}$ moments mark a 'defect-core' around Cr, which includes 4 V sites
with distance below 3\thinspace\AA $\;$ behaving quite differently from more
distant V sites. But importantly, there is also defect impact in a longer
range, e.g. also seen beyond 20 sites for a further-distant V atom along the
$c$-axis from the impurity site.
\begin{figure}[t]
\includegraphics[width=8cm]{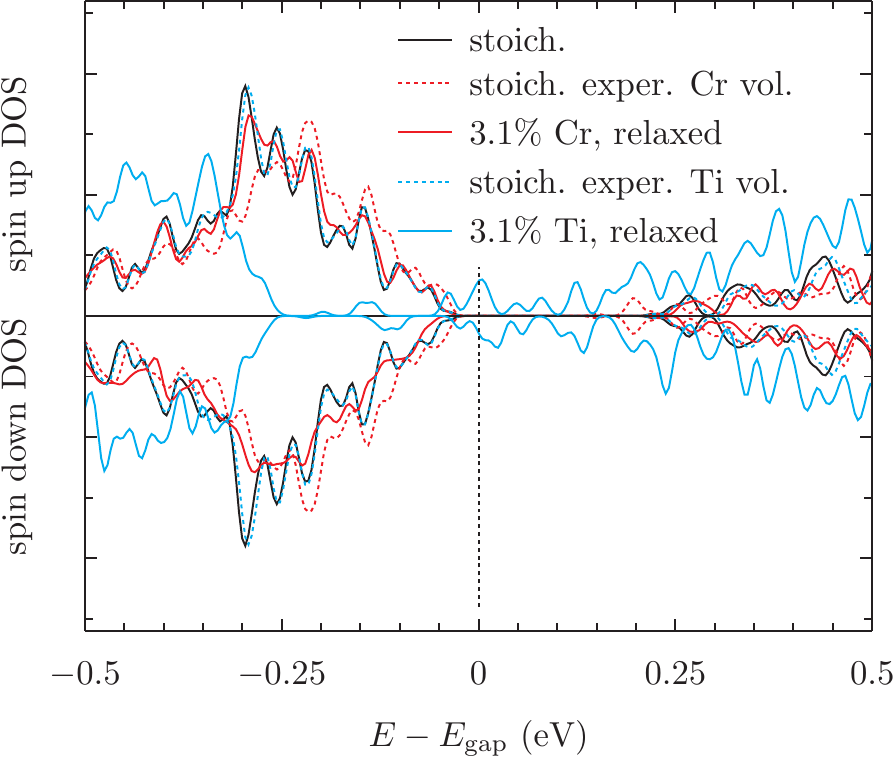}\\[0.2cm]
\caption{DFT+U density of states (with AF order) at 
$U_{\mathrm{eff}}=1.5$\thinspace eV. Shown are the pure stoichiometric system, 
the pure system at volumes equivalent to the experimental 2.8\% Cr and 3\% Ti 
doped volumes, and supercells with one atom of V substituted by Cr and Ti.}%
\label{fig:dftudos}%
\vspace*{0.4cm}
\includegraphics[width=8.5cm]{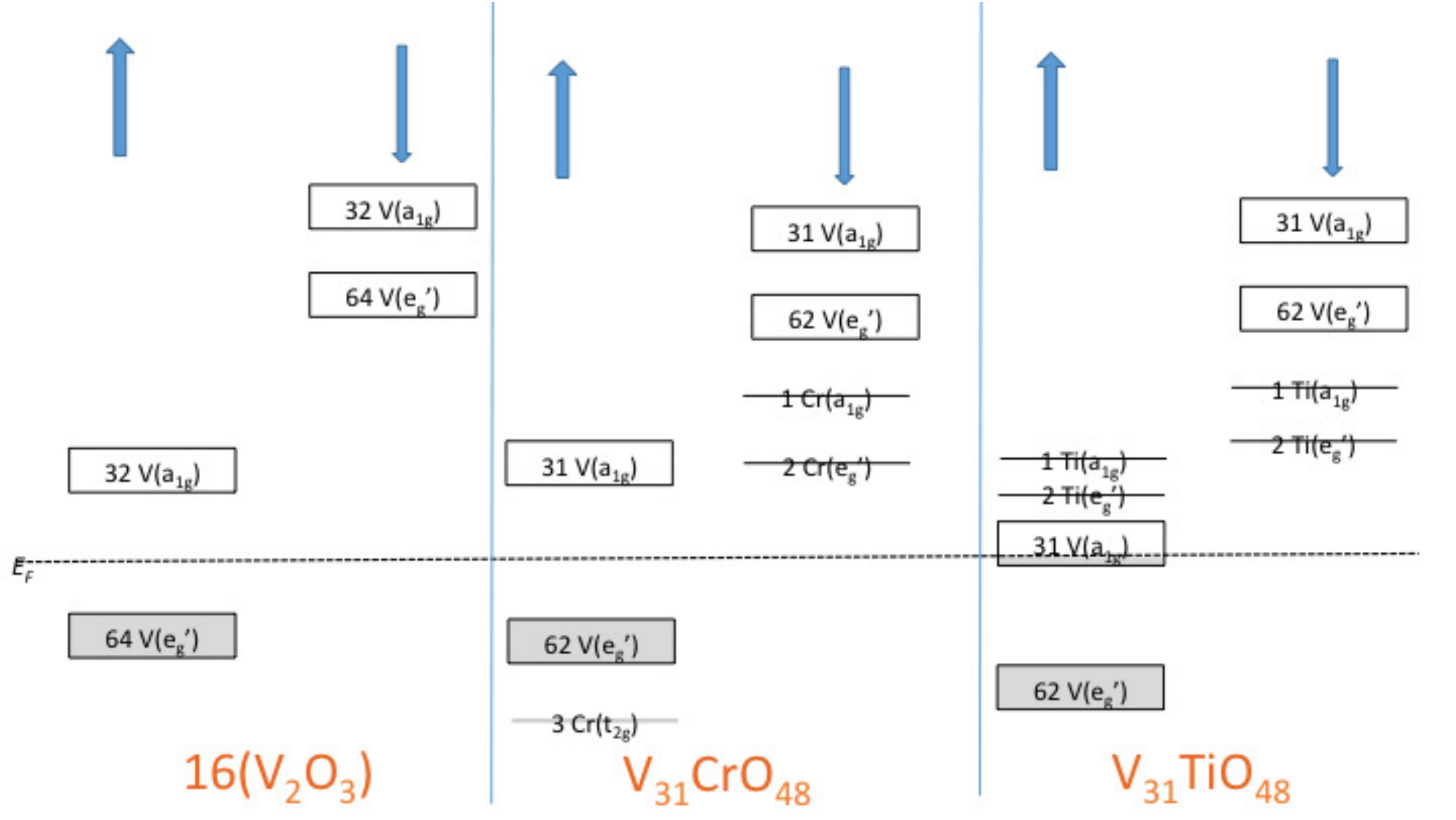}
\caption{DFT+U energy-level diagram for supercells of stoichiometric (left), 
relaxed Cr-doped (middle) and relaxed Ti-doped (right) V$_2$O$_3$.}%
\label{fig:dftulevels}%
\end{figure}

\textit{Hopping parameters}.--- The hopping parameters $t_{ij}^{mm^{\prime}%
}=\langle im|H_{\mathrm{loc}}^{(KS)}|jm^{\prime}\rangle$ between
transition-metal sites $ij$ and $t_{2g}$-like projected-local orbitals
$mm^{\prime}$ are a further source of information. Along the $c$-axis, as
expected, the $a_{1g}-a_{1g}$ hopping is dominant between nearest neighbors
(NN). It amounts to $t_{\mathrm{NN}}^{a_{1g}-a_{1g}}=\{-440,-358,-330,-374\}$%
\thinspace meV for stoichiometric, EA, unrelaxed Cr-doped and relaxed
Cr-doped. Here, in the latter two cases the Cr-V pair marks NN. The values are
in line with the associated lattice deformations, since Cr shifts towards V
with relaxation (see main text). Hoppings involving $e_{g}^{\pi}$ are at most
$\sim60$\thinspace meV along the $c$-axis, but the $e_{g}^{\pi}-a_{1g}$
hopping within the $ab$-plane takes on values $\sim200$\thinspace meV. Note
that an obvious hopping-driven mechanism underlying the metal-insulator
competition with Cr doping cannot be extracted.

\section{Supporting DFT+U information}
After the structural relaxation, the density of states may be calculated
within DFT+U for the small unit cells as well as for the 80-atom supercells.
The results for $U_{\mathrm{eff}}=1.5$\thinspace eV are displayed in
Fig.~\ref{fig:dftudos}.

A principle energy-level diagram (as discussed in the main text) originates from
those calculations for the different doping cases, and is pictured 
in Fig.~\ref{fig:dftulevels}.

\section{Supporting DFT+DMFT information}

\subsection{Orbital occupations and k-resolved spectral function}
In the main text, some orbital fillings were already given. The summarized
orbital $t_{2g}$ fillings for the different structural cases, and comparing
results from DFT and DFT+DMFT, are provided in Tab.~\ref{tab:t2gocc}.
\begin{table}[b]
\begin{ruledtabular}
\begin{tabular}{l|ll|ll}
&  \multicolumn{2}{c}{DFT} & \multicolumn{2}{c}{DFT+DMFT}\\
& $e_g^{\pi}$     & $a_{1g}$    &   $e_g^{\pi}$ & $a_{1g}$  \\  \hline
stoichiometric        & 1.44       &  0.56       &  1.58    & 0.42      \\
EA-Cr          & 1.44       &  0.56       &  1.60    & 0.40      \\[0.05cm]
3.1\% Cr unrel. & 1.42(2.40) &  0.55(0.82) &  1.87(2.67) & 0.10(1.28) \\
3.1\% Cr rel.         & 1.45(2.40) &  0.55(0.82) &  1.94(2.27) & 0.05(1.23) \\[0.05cm]
EA-Ti          & 1.44       &  0.56       &  1.60    & 0.40      \\
3.1\% Ti unrel. & 1.44(0.53) &  0.55(0.28) &  1.94(0.17)  & 0.09(0.11)   \\
3.1\% Ti rel.         & 1.43(0.45) &  0.55(0.26) &  1.70(0.09) & 0.33(0.06) \\
\end{tabular}
\end{ruledtabular}
\caption{Site-averaged V-$t_{2g}$ occupations in the different structural
cases, including EA volume and unrelaxed and relaxed explicitly
substituted supercells. The DFT+DMFT data is given for $T=193$~K. Values in
parentheses are for the dopant atom.}%
\label{tab:t2gocc}%
\end{table}
\begin{figure}[t]
\begin{center}
\includegraphics*[width=6.5cm]{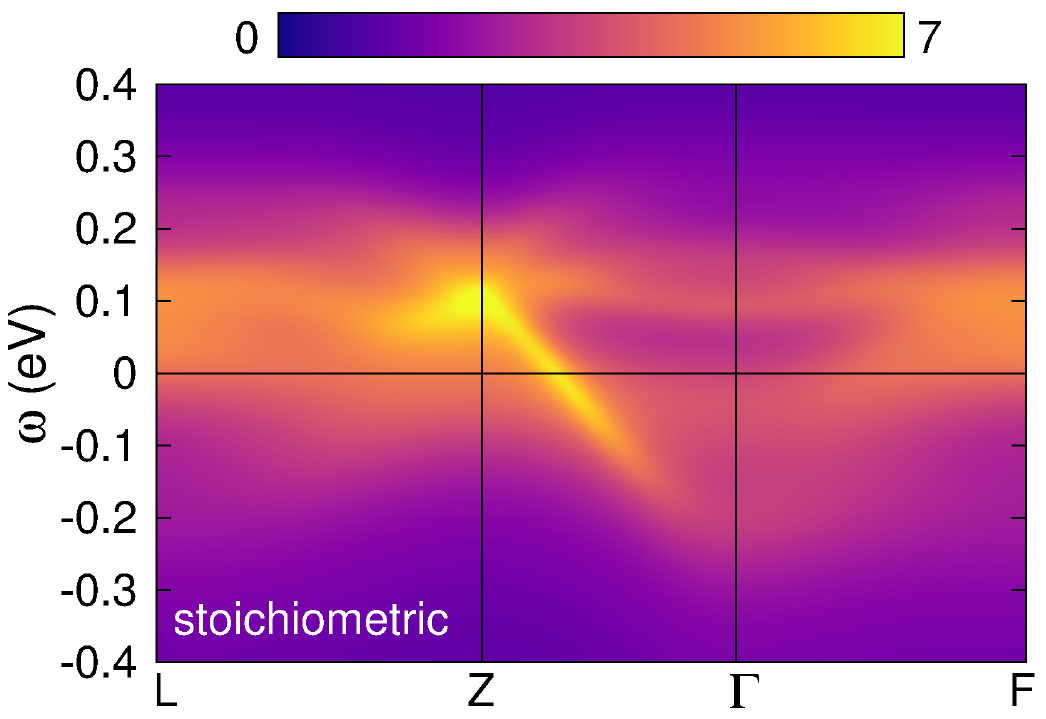}
\end{center}
\par
\includegraphics*[width=4.25cm]{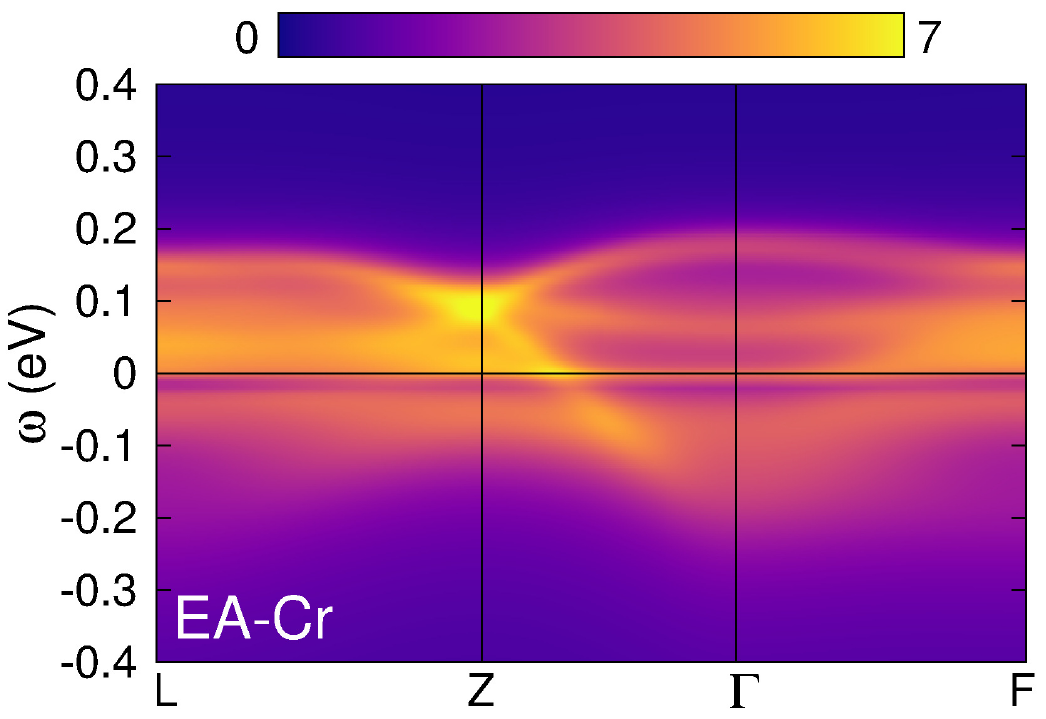}
\includegraphics*[width=4.25cm]{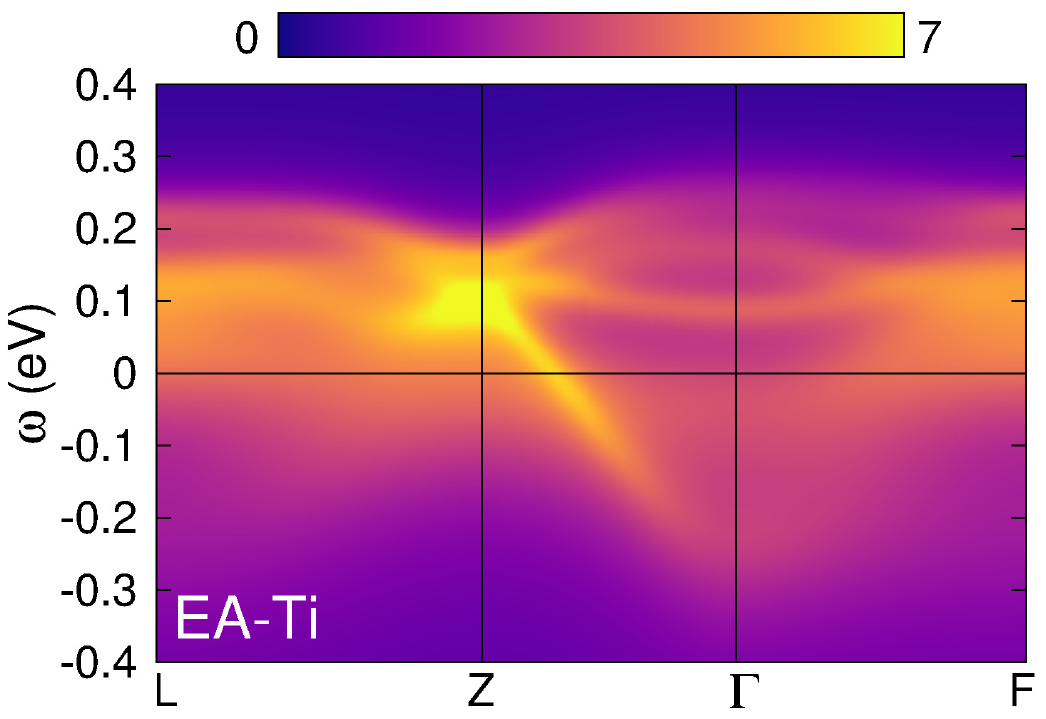} \caption{(Color online)
$k-$resolved DFT+DMFT low-energy spectral functions.
Top: stoichiometric V$_2$O$_3$. Bottom: EA-Cr structure (left) and
EA-Ti structure (right). All data at $T=193$\,K.}%
\label{fig:kres}%
\end{figure}
In order to further compare our treatment of the V$_{2}$O$_{3}$ compound with
experiment, the $k$-resolved spectral function $A(\mathbf{k},\omega)$ for the
stoichiometric corundum system~[16] is computed at $T=193$\,K and its
low-energy quasiparticle part is shown in the top part of Fig.~\ref{fig:kres}.
Strong renormalization and significant loss of coherency is observed even
close to the Fermi level. The weak electron pocket around $\Gamma$ above
$\sim-0.25$\,eV is in good agreement with recent angle-resolved
photoemission~[22], albeit the experimental energy depth below the
Fermi level is still larger with $\sim -0.4$\,eV. 

For comparison, we also show the $k$-resolved spectral function 
$A(\mathbf{k},\omega)$ for the EA-Cr/Ti
structures in the bottom of Fig.\ref{fig:kres}. Excpectedly, the low-energy
width is even stronger reduced for the Cr case, while for EA-Ti no obvious
qualitative difference to the stoichiometric case is visible 
But moreover interestingly, the EA-Cr case displays 
flat-band-like behavior close to $\varepsilon_{\rm F}^{\hfill}$, which is in 
line of our findings of the Cr-doped case begin close to a structural instability.

\subsection{Temperature effect in stoichiometric V$_{2}$O$_{3}$}
In addition to the metal-insulator transition (MIT) with Cr doping, the
V$_{2}$O$_{3}$ phase diagram displays MITs with temperature. In the present
context of this work, we focus on the high-$T$ part of the phase diagram and
do not deal with the transition to magnetic order at lower $T$. Then at
stoichiometry, there is a transition to a crossover region at around
$T=400$\,K beyond a critical point, where the differences between (incoherent)
metal and insulator disappear. The question arises if sole coherence-scale
arguments for the involved V-$t_{2g}$ orbital account for this high-$T$
behavior at stoichiometry.
\begin{figure}[t]
\includegraphics*[width=8.5cm]{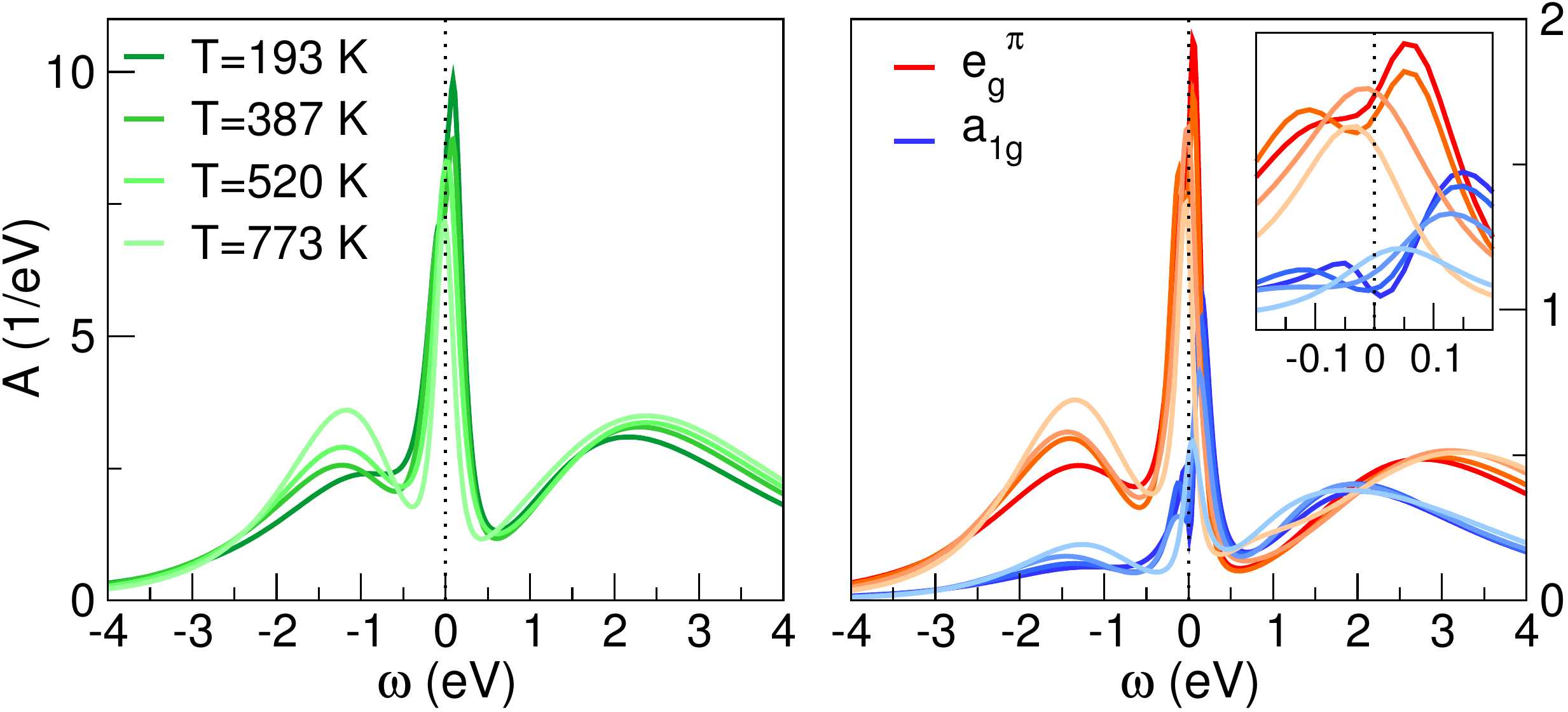}\caption{(Color online) Total
(left) and local (right) DFT+DMFT spectral function of stoichiometric 
V$_{2}$O$_{3}$ in the corundum phase with temperature. The color-intensity
changes for the local spectral functions correspond to the different
temperature scales given for the total functions.}%
\label{fig:stoicht}%
\end{figure}
\begin{figure}[b]
\includegraphics*[height=9cm]{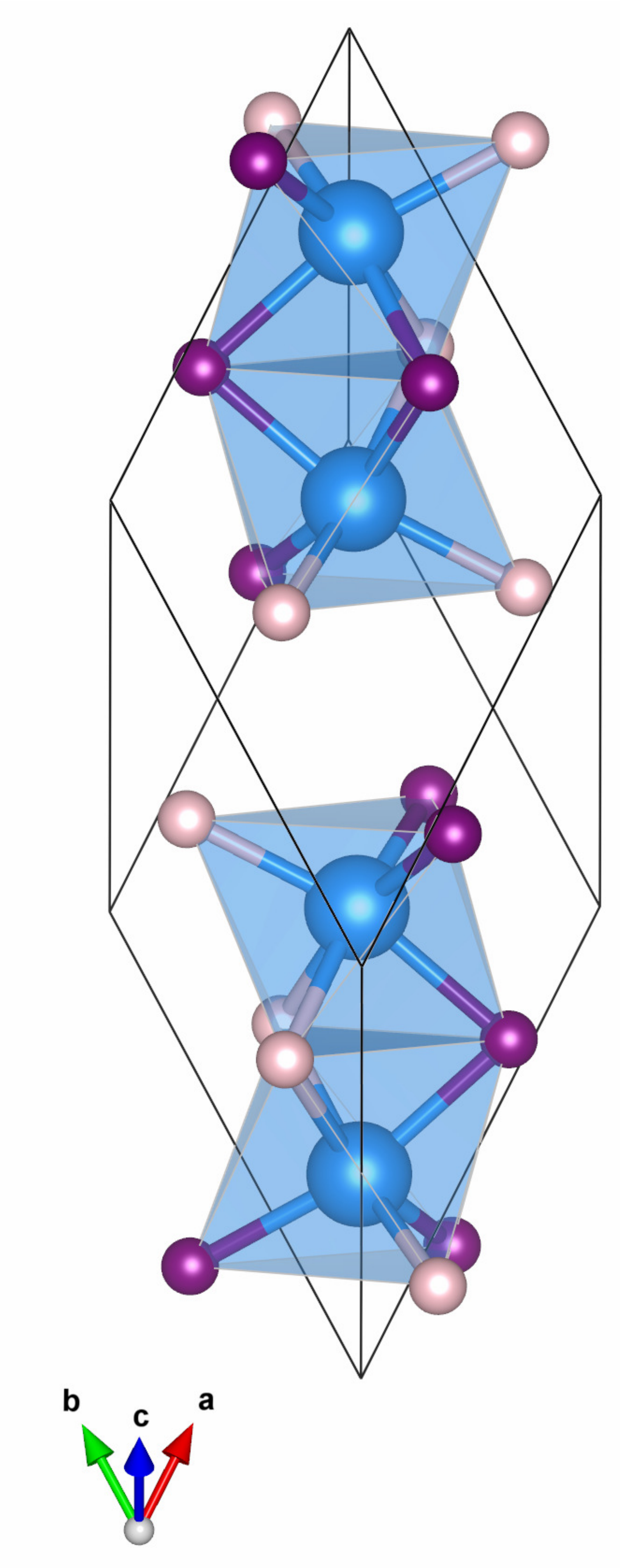}
\caption{(Color online) 
Basic two-formula-unit cell of the (quasi-)random V$_{2}$O$_{3}$ 
corundum structure with different pseudo-oxygen ions: $_{8-\delta}$O (violet), 
$_{8+\delta}$O (lightpink).}%
\label{fig:charge-struc}%
\end{figure}
Figure~\ref{fig:stoicht} shows that stoichiometric vanadium sesquioxide in the
ordered corundum phase remains metallic up to temperatures well above the
experimental transition temperature to the crossover regime. However
coherence-scale effects are visible, namely $T$-induced spectral weight
transfer from low energy to the Hubbard sidebands, and a reduction of the
QP-peak height at the Fermi level $\varepsilon_{\mathrm{F}}$. Right at
$\varepsilon_{\mathrm{F}}$, the hump structure disappears when entering the
experimental crossover region. Although for a detailed investigation of these
effects a $k$-dependent study is needed, it becomes obvious that a
(orbital-selective) localization of electrons due to complete loss of
coherence via heating is not taking place in the perfectly-ordered structure.
Rather, it is much more likely that the now $T$-induced creation of defects
such as vacancies, anti-site atoms or (dynamic) lattice distortions becomes
again relevant for the orbital-selective loss of coherence and the eventual
destruction of the metallic state.

\subsection{Charge-driven symmetry-breaking field}
For the investigation of symmetry breaking in V$_{2}$O$_{3}$ within a
standard-unit-cell approach via sole charge-density effects, we start from the
corundum crystal structure~[16]. The six oxygen basis
atoms are differentiated by random means in two groups: $_{8+\delta}$O and
$_{8-\delta}$O (cf. Fig.~\ref{fig:charge-struc}). Pseudopotentials with
effective nuclear charge $Z=8\pm\delta$ are constructed for these two groups
of atoms. Performing DFT+DMFT computations on this lattice corresponds to a
multi-site virtual-crystal-approximation (VCA) for symmetry-broken
defect-V$_{2}$O$_{3}$ without positional changes. Since our perturbation was
completely random, and we are only interested in qualitative effects, we did
not see any need to try other similar configurations.\\[0.5cm]

\section{References}

\noindent
[1] G. Kresse and J. Furthm\"uller, Phys. Rev. B {\bf 54}, 11169 (1996).\\
\noindent
[2] J. P. Perdew, K. Burke, and M. Ernzerhof, Phys. Rev. Lett. {\bf 77}, 3865 (1996).\\
\noindent
[3] V. I. Anisimov, I. V. Solovyev, M. A. Korotin, M. T. Czyzyk, and G. A. Sawatzky, 
Phys. Rev. B {\bf 48}, 16929 (1993).\\
\noindent
[4] S. G. Louie, K. M. Ho, and M. L. Cohen, Phys. Rev. B {\bf 19}, 1774 (1979).\\
\noindent
[5] B. Meyer, C. Els\"asser, F. Lechermann, and M. F\"ahnle,
Fortran 90 program for mixed-basis-pseudopotential calculations for crystals.\\
\noindent
[6] B. Amadon, F. Lechermann, A. Georges, F. Jollet, T. O.
Wehling, and A. I. Lichtenstein, Phys. Rev. B {\bf 77}, 205112 (2008).\\
\noindent
[7] V. I. Anisimov, D. E. Kondakov, A. V. Kozhevnikov,
I. A. Nekrasov, Z. V. Pchelkina, J. W. Allen, S.-K. Mo,
H.-D. Kim, P. Metcalf, S. Suga, et al., Phys. Rev. B {\bf 71}, 125119 (2005).\\
\noindent
[8] A. N. Rubtsov, V. V. Savkin, and A. I. Lichtenstein, Phys. Rev. B {\bf 72}, 035122 
(2005).\\
\noindent
[9] P. Werner, A. Comanac, L. de' Medici, M. Troyer, and
A. J. Millis, Phys. Rev. Lett. {\bf 97}, 076405 (2006).\\
\noindent
[10] O. Parcollet, M. Ferrero, T. Ayral, H. Hafermann,
I. Krivenko, L. Messio, and P. Seth, Comput. Phys. Commun. {\bf 196}, 398 (2015).\\
\noindent
[11] P. Seth, I. Krivenko, M. Ferrero, and O. Parcollet, Comput. Phys. Commun. {\bf 200}, 
274 (2016).\\
\noindent
[12] A. I. Poteryaev, J. M. Tomczak, S. Biermann, A. Georges, A. I. Lichtenstein, A. N. 
Rubtsov, T. Saha-Dasgupta, and O. K. Andersen, Phys. Rev. B {\bf 76}, 085127 (2007).\\
\noindent
[13] K. Held, G. Keller, V. Eyert, D. Vollhardt, and V. I. Anisimov, Phys. Rev. Lett. {\bf 86}, 
5345 (2001).\\
\noindent
[14] D. Grieger, C. Piefke, O. E. Peil, and F. Lechermann, Phys. Rev. B {\bf 86}, 155121 (2012).\\
\noindent
[15] I. Leonov, V. I. Anisimov, and D. Vollhardt, Phys. Rev. B {\bf 91}, 195115 (2015).\\
\noindent
[16] P. D. Dernier, J. Phys. Chem. Solids {\bf 31}, 2569 (1970).\\
\noindent
[17] P. D. Dernier and M. Marezio, Phys. Rev. B {\bf 2}, 3771 (1970).\\
\noindent
[18] S. Chen, J. E. Hahn, C. E. Rice, and W. R. Robinson, J. of Solid State Chem. {\bf 44}, 
192 (1982).\\
\noindent
[19] V. Eyert, Ann. Phys. (Leipzig) {\bf 11}, 650 (2002).\\
\noindent
[20] F. Cyrot-Lackmann, J. Phys. Chem. Solids {\bf 29}, 1235 (1968).\\
\noindent
[21] A. P. Sutton, Electronic Structure of Materials (Oxford University Press, 1993).\\
\noindent
[22] I. Lo Vecchio, J. D. Denlinger, O. Krupin, B. J. Kim, P. A. Metcalf, S. Lupi, 
J. W. Allen, and A. Lanzara, Phys. Rev. Lett. {\bf 117}, 166401 (2016).

\end{document}